\documentclass[11pt,preprint]{aastex}
\usepackage{graphicx}
\usepackage{natbib}
\usepackage{caption}

\shorttitle{Radio continuum observations of local star-forming galaxies}
\shortauthors{Rabidoux et al.}

\title{Radio continuum observations of local star-forming galaxies
  using the Caltech Continuum Backend on the Green Bank Telescope}

\author{Katie Rabidoux, D.J. Pisano\altaffilmark{1}} 
\affil{Department
  of Physics and Astronomy, West Virginia University, 135 Willey St., P.O. Box 6315
  Morgantown, WV 26506} 
\altaffiltext{1}{Adjunct Assistant Astronomer at National Radio Astronomy Observatory, P.O. Box 2, Rt. 28/92, Green Bank, WV 24944}

\author{Amanda A. Kepley\altaffilmark{2,3}, Kelsey E. Johnson\altaffilmark{4}}
\affil{Department of Astronomy, University of Virginia, P.O. Box 400325, Charlottesville, VA 22904}
\altaffiltext{2}{Current address: National Radio Astronomy Observatory, P.O. Box 2, Rt. 28/92, Green Bank, WV 24944}
\altaffiltext{3}{Visiting Research Associate at National Radio Astronomy Observatory, 520 Edgemont Road, Charlottesville, VA 22904}
\altaffiltext{4}{Adjunct Assistant Astronomer at National Radio Astronomy Observatory, 520 Edgemont Road, Charlottesville, VA 22904}

\and
\author{Dana S. Balser}
\affil{National Radio Astronomy Observatory, 520 Edgemont Road, Charlottesville, VA 22904}

\begin{document}

\begin{abstract}

We observed radio continuum emission in 27 local (D $<$ 70 Mpc)
star-forming galaxies with the Robert C. Byrd Green Bank Telescope
between 26 GHz and 40 GHz using the Caltech Continuum Backend. We
obtained detections for 22 of these galaxies at all four sub-bands and
four more marginal detections by taking the average flux across the
entire bandwidth. This is the first detection (full or marginal) at
these frequencies for 22 of these galaxies. We fit spectral energy
distributions (SEDs) for all of the four-sub-band detections. For 14
of the galaxies, SEDs were best fit by a combination of thermal
free-free and nonthermal synchrotron components. Eight galaxies with
four-sub-band detections had steep spectra that were only fit by a
single nonthermal component. Using these fits, we calculated supernova
rates, total number of equivalent O stars, and star formation rates
within each $\sim 23\arcsec$ beam. For unresolved galaxies, these
physical properties characterize the galaxies' recent star formation
on a global scale. We confirm that the radio-far-infrared correlation
holds for the unresolved galaxies' total 33 GHz flux regardless of
their thermal fractions, though the scatter on this correlation is
larger than that at 1.4 GHz. In addition, we found that for the
unresolved galaxies, there is an inverse relationship between the
ratio of 33 GHz flux to total far-infrared flux and the steepness of
the galaxy's spectral index between 1.4 GHz and 33 GHz. This
relationship could be an indicator of the timescale of the observed
episode of star formation.
\end{abstract}


\section{Introduction}
Radio continuum emission traces star formation on timescales of up to
100 Myr \citep{C92}. Two physical processes resulting from massive
star formation produce most of the radio continuum emission between 1
and 100 GHz in star-forming galaxies: (1) nonthermal synchrotron
emission from relativistic electrons accelerated by magnetic fields as
a result of recent supernovae and (2) thermal free-free emission from
gas ionized by young massive stars \citep{C92}. The nonthermal
emission is closely tied to the number of supernova-generating massive
stars produced in recent episodes of star formation, while the thermal
emission gives a nearly direct measure of the current equivalent
number of O stars via the ionizing flux in the sampled area. Since
each component traces a physical process with a well-known timescale,
we can use measurements of the radio continuum to determine star
formation rates and constrain the ages of recent episodes of star
formation.

Recent studies of nearby star-forming galaxies with interferometers
have emphasized resolving individual star-forming regions
\citep*[e.g.][]{B00,J03,JK03,J04,T06,R08a,J09,A11}. Since radio
continuum emission is not affected by extinction, it can be used to
observe deeply embedded regions of current star formation that have
not yet shed their surrounding material and are thus invisible at
shorter wavelengths. These studies have taken advantage of
interferometers' exceptional spatial resolution to probe very young
starbursts whose optical emission is obscured by dust. While these
studies have been invaluable for determining star formation properties
in galaxies outside of our own, the high angular resolution and
missing short-spacing data of interferometers, especially at higher
frequencies, ``resolves out'' the diffuse radio continuum emission
outside of compact star-forming regions. This effect disproportionally
impacts synchrotron emission, which tends to be much more diffuse than
the primarily thermal emission surrounding areas of ongoing massive
star formation \citep{J09}. Unlike interferometers, single dish
telescopes are not plagued by missing short spacings. Therefore, these
telescopes provide a way to simultaneously measure the compact thermal
and diffuse non-thermal components of a galaxy's radio continuum
emission in order to characterize its \emph{global} star formation
properties.

Determining the relative contributions of the thermal and nonthermal
components of the measured flux of entire galaxies can be challenging.
Fortunately, each component has a distinct behavior with respect to
frequency, and therefore we can model radio continuum emission with a
simple two-component fit. Radio continuum flux follows a power law
relation such that $S_{\nu} \propto \nu^{\alpha}$, where $\alpha$ is
the spectral index that is characteristic of a source's
emission. Optically thin thermal emission exhibits $\alpha =$ -0.1,
and nonthermal emission exhibits $\alpha \approx$ -0.8 \citep{C92}. To
determine a reliable fit to these parameters, observations sampling
the same physical area at multiple, widely-spaced frequencies are
required. If only one frequency is observed, it is impossible to
determine the relative contributions of each emission process without
previous knowledge of the source. Since single dish telescopes are
sensitive to both compact thermal and diffuse nonthermal emission over
large spatial extents, they are useful for constraining the
large-scale properties of multiple components of a galaxy's radio
continuum emission, and are thus powerful probes of star formation.

The goal of this paper is to characterize the \emph{global} star
formation properties of local galaxies. Our observations were taken in
four independent channels continuous in frequency across the full
26-40 GHz span of Ka band. This range in frequencies is where a
typical star-forming galaxy's global radio continuum emission would be
expected to contain relatively equal amounts of flux from
steep-spectrum synchrotron and flat-spectrum thermal sources
\citep{C92}. Our observations are thus ideal for approximating the
relative contributions of each type of emission at this ``lever arm''
frequency range. Using new radio continuum observations centered at
27.75 GHz, 31.25 GHz, 34.75 GHz, and 38.25 GHz, as well as archival
NVSS 1.4 GHz and IRAS 60 $\rm \mu m$ and 100 $\rm \mu m$ data, we have
determined these galaxies' thermal fractions and star formation
rates. We have also explored the radio-far-infrared correlation in
these galaxies and its implications for their star formation
timescales. We will describe the galaxy sample and our observations
and data reduction in Section 2, present our results and address the
process of fitting spectral energy distributions to our data in
Section 3, and finally conclude in Section 4.

\section{Data}

\subsection{Sample Selection}

\begin{figure}
\begin{center}
\includegraphics[angle=270,width=1.0\linewidth]{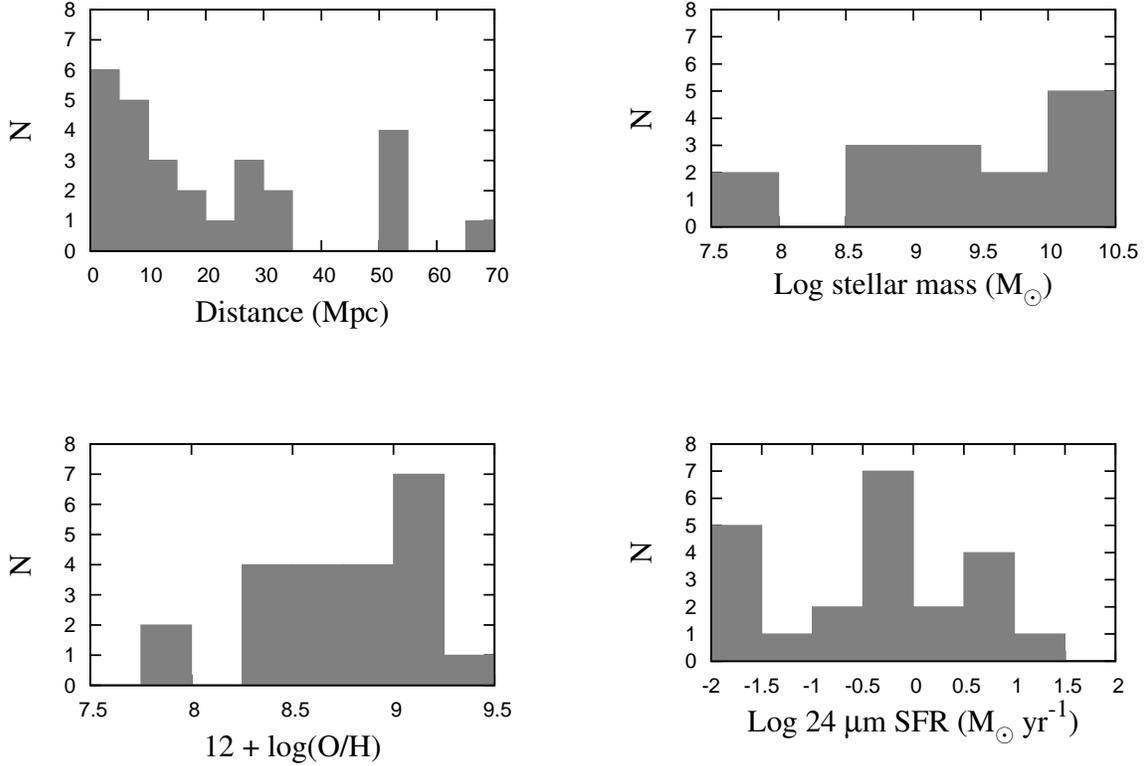}
\end{center}
\caption{\emph{Top Left:} Distribution of distances in our sample. \emph{Top Right:} Distribution of K band optical galactic stellar masses in our sample estimated using the (B-V) color and the expression in \citet{BJ01}. \emph{Bottom Left:} Metallicity distribution for our sample estimated from the B band absolute magnitude using the expression in \citet{T04}. \emph{Bottom Right:} Star formation rate for galaxies in our sample calculated using the 25$\mu$m IRAS fluxes for our sample and the expression in \citet{Cal10}. Not all galaxies are represented in every histogram.}
\label{fig:sampleproperties}
\end{figure}

We selected a heterogeneous group of 27 local (D $<$ 70 Mpc),
well-studied star-forming galaxies with known thermal radio continuum
emission. Our sample contains galaxies spanning a variety of shapes,
sizes, and environments, from blue compact dwarfs to grand-design
spirals, including major and minor mergers, with members of compact
groups as well as more isolated galaxies (see Table
\ref{fig:obssummary} for galaxy types). The intention was to observe
as many types of star-forming galaxies as possible to probe star
formation in a diverse range of environments. See Table
\ref{fig:obssummary} and Figure \ref{fig:sampleproperties} for sample
properties. For more information on each galaxy's previous radio
continuum observations and discussions of their properties, see the
papers named in Table \ref{fig:obssummary}.

The galaxies in our sample span a range of distances (1-70 Mpc) and
properties. They all have previously detected radio emission and
ongoing star formation that covers three orders of magnitude in star
formation rate. Thus, they are strong targets for a study of global
radio continuum properties at a frequency range that probes both
thermal free-free and nonthermal synchrotron star formation
indicators. As seen in Figure \ref{fig:sampleproperties}, these
galaxies are largely less massive and have higher star formation rates
than the Milky Way, and have subsolar metallicities. However, their
properties are not so similar that they can be considered as a single
class. It would not be surprising if their radio continuum properties
also encompassed a range of values. Our analysis is best understood as
reflecting properties of nearby star-forming galaxies, though it is
beyond the scope of this paper to perform detailed analysis on each
galaxy individually.

\subsection{Observations and Data Reduction}
We observed the galaxies in our sample with the Caltech Continuum
Backend (CCB) on the Robert C. Byrd Green Bank Telescope
(GBT)\footnotemark{} using single pointings. The CCB is designed for
the GBT's dual-beam Ka band receiver spanning the entire range of
frequencies from 26-40 GHz. The primary observing mode of the CCB is a
70 second ``nod'', where each beam takes a turn as the on-source beam
while the other beam is off-source. We observed 24 of the galaxies
using a single nod each, while we observed eight galaxies using
multiple nods, which we then averaged. Five galaxies in our sample
were observed on two different nights with both of these methods; we
treated these on a case-by-case basis and chose the observation(s)
with the best weather and elevation conditions. We used the standard
NRAO primary flux calibrators 3C 147 and 3C 48 for flux calibration,
as well as nearby pointing calibrators to ensure accurate
pointing. See Table \ref{fig:obssummary} for a summary of the
observations.

\footnotetext{The National Radio Astronomy Observatory is a facility
  of the National Science Foundation operated under cooperative
  agreement by Associated Universities, Inc.}

We reduced our data using IDL reduction routines developed by B.
Mason \citep[for details on the data reduction process,
  see][]{M09}. Data with wind speeds over 5 $\rm m \ s^{-1}$ were
excluded due to the possibility of large pointing errors. We detected
22 galaxies in all four channels. When a galaxy's flux was lower than
the 2$\sigma$ level in one or more of the four channels, we combined
the channels' fluxes to produce one average flux across the entire
band, centered at 33 GHz. One galaxy, Pox 4, was detected at the
$5\sigma$ level after averaging the four channels. Three additional
galaxies were marginally detected (between 2$\sigma$ and 3$\sigma$)
using this method. We report an upper limit 33 GHz flux for one
galaxy, Tol 35. The galaxies' observed fluxes are reported in Table
\ref{fig:uncorrectedflux}.

Since the angular size of a telescope's beam is inversely proportional
to the frequency observed, the beam size of the GBT varies appreciably
across the 26-40 GHz range of our observations ($\sim27\arcsec$ in the
lowest-frequency sub-band versus $\sim19\arcsec$ in the
highest-frequency sub-band, see Figure ~\ref{fig:beamsizes} for an
illustration). We followed the procedure of \citet{M10} to correct for
differing beam sizes in each of the four sub-bands. First, we imaged
an archival VLA radio continuum map of each galaxy (typically at
frequencies of 4-10 GHz) using the AIPS task IMAGR. For these maps, we
selected the archival UV data from the NRAO science data
archive\footnotemark{} with the closest beamsize to our Ka-band
data. We explicitly imposed each of the four CCB beam sizes on these
images using BMIN and BMAJ (assuming a circular beam). We determined
correction factors for each beam by normalizing the flux contained in
each beam area in the archival map to the flux in the 34.75 GHz port's
$\sim21\arcsec$ beam. This procedure partially adjusts for more flux
being observed at lower frequencies due to these frequencies'
intrinsically larger beam sizes. We then could approximate
``beam-matched'' flux measurements to determine spectral indices
between 26-40 GHz (see Figure \ref{fig:histogram} for an illustration
of the galaxies' spectral indices before and after applying the
corrections). NGC 1222 did not have available archival data, so we
applied to it the average correction factors of all of the other
galaxies. We emphasize that these correction factors are only
approximate. In many cases, they are based on resolved archival images
that may not contain all of the galaxies' radio flux. In particular,
these resolved images may contain most of the thermal emission, which
tends to be compact, but underestimate the galaxies' nonthermal
emission, which tends to be diffuse. This could bias the correction
factors to be closer to 1.0 than should be the case, especially in the
highest-frequency (and thus highest-resolution) channel. See Table
\ref{fig:beamcorrections} for the beam correction factors for each
galaxy. 

The dominant sources of uncertainty in our beam corrections are
systematic errors due to the geometries of our sources. The smallest
corrections possible are for a galaxy whose most diffuse, extended
flux is still contained within the smallest beam and is unresolved by
the lower-frequency interferometric observations. This type of source
would look identical to all four of the GBT beam sizes. In this case,
the correction factors would be 1.0 for each sub-band. For a source
much more extended than the beam sizes, the maximum deviations from no
beam corrections in each sub-band are -36$\%$, -19$\%$, 0$\%$, and
+21$\%$. Pointing offsets from the peak of radio emission can also be
sources of systematic error, though the errors depend on whether the
source is compact or extended, and the magnitude of the pointing
offset from the central peak of radio emission. These errors are
typically smaller than the maximum deviations discussed above. Since
we do not know how much diffuse emission is missing in the archival
data, we do not have enough information to quantify uncertainties in
the beam correction factors for each galaxy.

\footnotetext{https://archive.nrao.edu}

Many of the galaxies that we observed were more extended than the
$\sim 23\arcsec$ beam size of the GBT at 33 GHz. In these cases, radio
continuum fluxes and star formation rates should only be interpreted
as covering the inner $\sim 23\arcsec$ of the galaxies. The galaxies
with resolved lower-frequency archival data that was more extended
than the beam size are flagged with a ``1'' in Table
\ref{fig:correctedflux}.

The CCB has a beam separation of $78\arcsec$ between the ``on'' and
``off'' beams. M 51 and M 101 are more extended than that separation
in both optical images (see Figure \ref{fig:beamseparation}) and in
maps of their lower-frequency radio continuum emission
\citep{K84,G90}. In these cases, our flux measurements may be lower
than the true amount of flux contained within the beam. There is
likely to be radio continuum emission at the ``off'' positions, which
would cause an oversubtraction of flux in the reduction process.

\section{Results and Discussion}

\subsection{Fluxes}
For 22 of the 27 galaxies that we observed, the fluxes we report are
the first detections (either in all four sub-bands or averaged) at
$\sim$33 GHz. Four of the galaxies in our sample were previously
observed with the CCB by \citet{M12}, three of which were detected (M
101 was reported as an upper limit by \citet{M12} and is a $2.6\sigma$
marginal detection when the four sub-bands were averaged in our
observations). Only one galaxy in our sample, Tol 35, was not detected
when averaging four sub-bands' fluxes. Its $3\sigma$ upper-limit flux
is 0.87 mJy. This galaxy was observed at a very low elevation
($7.9^{\circ}$), so it was observed with large atmospheric
extinction. See Table \ref{fig:uncorrectedflux} for the uncorrected
fluxes, and Table \ref{fig:correctedflux} for the fluxes corrected for
the differing beam sizes of each frequency.

\subsection{Spectral energy distribution fitting}
We fit a spectral energy distribution (SED) for each galaxy that was
detected in all four sub-bands using the four CCB fluxes and archival
NRAO VLA Sky Survey (NVSS) 1.4 GHz fluxes (measured with a 45$\arcsec$
aperture). We assumed a two-component fit of nonthermal emission with
a spectral index $\alpha_{N}$ = -0.8 and thermal emission with a
spectral index $\alpha_{T}$ = -0.1. These fits are plotted in Figure
\ref{fig:sedpanel}. Though the spectral index of nonthermal emission
can vary (this phenomenon is described further in Section 3.2.1), we
used this simple model because we only fit to five data points for
each galaxy; our model did not include enough data to justify
additional free parameters. We do not see evidence of anomalous dust
emission in the observed regions of these galaxies \citep[for
  explanations of anomalous dust, see][]{D98,M10}. Our observations
are also at frequencies low enough to have negligible contributions
from the low-frequency tail of the dust blackbody. Therefore, we did
not include any thermal dust emission in our fits. Our spectra also do
not show the inverted structure characteristic of self-absorption or
optically thick thermal emission, so we did not include either of
these components. From these fits, we determined each galaxy's thermal
and nonthermal fluxes at 33 GHz.

None of the galaxies have globally flat spectra indicative of purely
thermal emission, nor the inverted spectra seen in some resolved
observations of very young, obscured thermal sources. Thermal emission
was the primary component at 33 GHz in some galaxies, while others had
less prominent or even negligible thermal components in the observed
regions. In contrast to radio continuum studies done at high spatial
resolution, our single dish observations detect the diffuse
synchrotron emission produced by past supernovae in addition to the
strong compact thermal emission from H II regions, so the spectral
indices that we derive are typically much steeper than those derived
only from detections of compact radio sources. Since our observations
do not spatially separate regions of thermal and nonthermal emission,
we cannot further distinguish the two components in that way.

\subsubsection{Galaxies with steep radio spectra}
The fitted spectra for eight of the 27 galaxies (Arp 217, NGC 4449,
NGC 2903, Maffei II, NGC 4038, M 51, NGC 4490, and NGC 1741) are
significantly steeper than can be fit by a combination of thermal
($\alpha_{T}$ = -0.1) and nonthermal ($\alpha_{N}$ = -0.8) components
(see Figure \ref{fig:sedpanel}). When we could not fit a galaxy's SED
with both the thermal and nonthermal components at the $2\sigma$
level, we used only a single-component fit that assumed no thermal
flux and a fixed nonthermal spectral index of $\alpha_{N}$ = -0.8 for
consistency. The thermal fluxes and associated properties of this
group of galaxies are reported as upper limits. We used the total flux
in the 34.75 GHz channel plus 3$\sigma$ as a conservative upper limit
to the thermal flux in these cases.

There are two possible explanations for the steep spectra that we see
in some galaxies. There could be technical considerations due to
imperfect beam-matching in our data, or there could be physical
processes taking place within these galaxies causing their spectra to
steepen at high frequencies. In order to have more accurate SED
fits---and more precise star formation rates---we would need to have
beam-matched observations of the same regions at many different
frequencies. 

The correction factors for differing beam sizes that are given in
Table \ref{fig:beamcorrections} are limited by being calculated using
higher-resolution data that could be missing extended emission. If
extended emission is missing in the archival data, the correction
factors in Table \ref{fig:beamcorrections} could be closer to 1.0 than
is actually the case. While all of the correction factors calculated
act to flatten the SED between 26 GHz and 40 GHz with respect to the
uncorrected data, it is possible that they do not flatten the SED
enough if they do not reflect contributions from extended emission (as
discussed in Section 2.2). In addition, we did not correct for
mismatched beams between the $\sim 45\arcsec$ NVSS data and the $\sim
23\arcsec$ CCB data. This beam difference only affects resolved
galaxies (those marked with a ``1'' in Table \ref{fig:correctedflux}),
which comprise $33\%$ of our sample. It is possible that synchrotron
emission is more adversely affected by the differences in beam sizes
than thermal emission. More diffuse synchrotron emission could be
undetected at higher frequencies (and thus smaller beam sizes) than
would be expected for a smooth flux distribution observed with two
apertures of different sizes. If this is the case in our observed
regions, it could explain why some of our galaxies' spectra steepen at
the frequencies we observed. It is also possible that the choice of
where the GBT beams were pointed within a galaxy could affect its
fluxes in different beam sizes. If the beam is not centered on the
galaxy (in the case of unresolved galaxies) or is not centered on a
bright knot of emission (in the case of resolved galaxies), the
smaller beams could contain even less flux than would be expected
after corretions for the beams' areas. NGC 1741 and NGC 4490 are
likely affected by pointing offsets, as seen by comparing the GBT
pointing in Table \ref{fig:obssummary} to previous radio continuum
maps in Figure 2 of \citet{B00} and Figure 4 of \citet{A11}. As
described in Section 2.2, pointing offsets from the peak of radio
continuum emission result in the need for larger beam correction
factors than derived from the archival radio continuum data, the lack
of which result in steep spectra at the observed frequencies.

In addition to the technical issue of mismatched beam sizes, there are
possible physical explanations for steep spectra in star-forming
galaxies. It is difficult to distinguish between a spectrum with a
nonthermal component having $\alpha_{N} \approx -0.8$ coupled with a
low thermal fraction from a spectrum with a steeper nonthermal
component coupled with a relatively high thermal fraction
\citep{C92}. Though the spectral indices that we used are typical
values \citep{C92}, they can vary depending on the physical parameters
of the observed regions. Thermal emission can have a positive spectral
index if the emission regions are optically thick, though we do not
see any evidence that this is occuring on the angular scale of our
observations. Nonthermal spectral indices can be positive at low
frequencies due to synchrotron self-absorption (which we do not
observe), or become more negative with increasing frequency and
increasing scale height from the disk due to aging cosmic ray
electrons losing energy as they propagate outward from their parent
supernovae \citep{S85,CK91,H09}. \citet{K11b} calculated the timescale
for synchrotron losses for cosmic ray electrons in NGC 4214 to be 44
Myr at 1.4 GHz and 18 Myr at 8.5 GHz. There is also some evidence of
steepening spectra at higher frequencies ($\gtrsim$ 10 GHz) for
luminous and ultra-luminous infrared galaxies, as well as in the
post-starburst galaxy NGC 1569 \citep{I88,L04,C08,C10,L11}. These
authors hypothesize that winds or outflows may disperse synchrotron
emission from its parent source more quickly than would be expected
for simple diffusion. This rapid dispersal could cause a dearth of
synchrotron emission at higher frequencies on shorter timescales than
would be predicted from the timescale of energy loss. \citet{L11} also
hypothesize that there could be a modified injection spectrum in
galaxies where this is the case. Our sample of galaxies does not
contain any LIRGs or ULIRGs, and we do not see steepening in our
measurements of NGC 1569. We are only observing the inner region of
NGC 1569, while the dispersed synchrotron emission resides in its
outer halo, so it is not surprising that we do not observe a
steepening spectrum in this galaxy.

We suspect that the steep spectra seen in our sample are primarily a
result of imperfect beam matching as discussed above. This is
especially likely to be the case for the galaxies resolved by the GBT
at 33 GHz, since these galaxies will have emission that is outside of
the view of the GBT beam but is included in the NVSS flux. As
discussed earlier, the galaxies that appear unresolved in archival
maps could still have diffuse synchrotron emission that was not
detected in the archival data but that is more extended than the
$23\arcsec$ beam at 33 GHz. Five of the eight resolved galaxies that
were detected in all sub-bands had steep spectra (four out of those
five are classified as spiral galaxies), while only three of the
fourteen unresolved galaxies had this feature. Two of these three are
classified as SABbc galaxies, while the third is classified as
peculiar. It is possible that these steep-spectrum galaxies contain
emission in their spiral arms that is extended with respect to the
GBT's smaller beam but is observed in the NVSS data. In Figure
\ref{fig:sedpanel}, most of the galaxies with steep spectra (and thus
single component fits) also showed the NVSS 1.4 GHz data point being
located above the best-fit line expected for purely nonthermal
emission. This could be a consequence of the larger beam at 1.4 GHz
sampling a larger physical area of emission. Even so, the alternative
physical explanations merit consideration, especially in the case of
the unresolved galaxies. In Figure \ref{fig:rad-IR}, which will be
discussed further in Section 4.6, the three unresolved galaxies with
steep spectra (NGC 1741, NGC 2903, and Arp 217) have elevated 1.4 GHz
fluxes with respect to what would be expected from the radio-far
infrared correlation. Since the 33 GHz fluxes of these galaxies are
not similarly elevated with respect to their far-infrared fluxes in
Figure \ref{fig:rad-IR}, their steep radio spectra may indicate an
internal physical process that strongly increases the amount of
synchrotron emission.

\subsection{Thermal fractions}
The average thermal fraction fit by two-component models at 33 GHz was
54$\%$, with a $1\sigma$ scatter of 24$\%$ and a range of
10$\%$-90$\%$. The average is consistent, albeit with large scatter,
with the average global thermal fraction at 33 GHz in star-forming
galaxies without active galactic nuclei following the relation
\begin{equation}
\frac{S}{S_{T}} \sim 1 + 10\left(\frac{\nu}{GHz}\right)^{0.1+\alpha_{N}}
\end{equation}
 where $\alpha_{N} = -0.8$ is the nonthermal spectral index, $S_{T}$
 is the thermal flux at a given frequency, and S is the total flux at
 that frequency \citep{CY90}. When a two-component fit was not
 possible, we report the thermal flux as the corrected flux at 34.75
 GHz plus $3\sigma$, which gives a very conservative upper limit. We
 expect from the galaxies' SEDs that their true thermal fractions are
 very low at 33 GHz, which we assume in the rest of our analysis.

\subsubsection{Implications for star formation timescales}
The large scatter in the thermal fraction is likely a consequence of
our heterogeneous galaxy sample; these galaxies are at different
stages of evolution and have different star formation rates, stellar
populations, and physical properties \citep{B00}. Some of them may
have a very recent ($<$ 10 Myr) burst of star formation that produces
a large amount of free-free emission that dominates their spectra from
1-100 GHz. Others may be in between episodes of very active star
formation and instead be experiencing a more quiescent phase, which
would result in a relatively low thermal fraction and steepening
nonthermal component at 33 GHz due to synchrotron energy losses at
high frequencies.

Thermal emission traces very recent star formation, since it comes
from ionized regions around short-lived, massive stars. For a single
starburst, a spectrum showing solely thermal emission requires that
too few supernovae have yet occurred to detect their emission. This
would constrain the starburst to be less than $\sim$ 6 Myr old (or
even younger, depending on the mass and lifetime of the most massive O
stars in the starburst; \citet{M89} find the lifetime of a 120
$M_{\sun}$ star to be 3.4 Myr). A complete absence of thermal flux
implies the absence of enough massive O stars to have detectable
free-free emission for a long enough period of time that the emission
has dissipated from its parent region. If this was the case, the
starburst is likely at least 30 Myr old (the lifetime of the least
massive supernova progenitors). On the other hand, nonthermal emission
probes star formation on longer timescales (30 Myr $< \tau <$ 100
Myr). It is produced by recent supernovae of stars that can be less
massive and have longer lifetimes than the O stars that produce
thermal emission \citep[see Figure 9 of][]{C92}. The presence of
nonthermal emission implies that the starburst is at least 6 Myr old
but younger than the timescale dictated by synchrotron energy loss for
the galaxy's magnetic field ($\sim$ 100 Myr) \citep{C92}.

We note that there are limits to the amount of each component that we
can detect, so the timescales quoted in the previous paragraph are
only approximate. To constrain how much nonthermal emission could be
present in a spectrum that appears purely thermal, we generated
spectra with varying thermal fractions with fluxes at the same five
frequencies as those in our data set (1.4 GHz, 27.75 GHz, 31.25 GHz,
34.75 GHz, and 38.25 GHz) and 10$\%$ errors on the fluxes. When these
spectra are fit with a two-component model assuming $\alpha_{T} =
-0.1$ and $\alpha_{N} = -0.8$, nonthermal emission can only be
detected in the spectra for thermal fractions less than 97$\%$. This
means that the galaxy could have some nonthermal emission (up to $3\%$
for $10\%$ errors on the fluxes), but the emission would be
undetectable and thus the starburst would appear younger than it
is. Similarly, a spectrum could look like it contains no thermal
emission while actually containing quite a bit. For the same spectra
with $10\%$ errors on the fluxes, thermal fractions of up to $20\%$
resulted in undetectable thermal components. This means that a galaxy
could look like its massive star formation has ceased while still
having a small thermal component.

For the galaxies in our sample, this picture could be more
complicated. The quoted timescales in this section are for an isolated
single starburst. Since our observations measure star formation
properties on large angular scales, the galaxies may have multiple
overlapping generations of star formation that are not easily
separated in time. We are also sampling different structures and
physical scales in each galaxy. For some galaxies, we are only
observing the most central region. For these galaxies, we may be
missing the majority of the ongoing star formation happening in outer
regions and spiral arms. For the more compact galaxies, however, we
are likely measuring the entirety of the galaxy's star formation
within the GBT beam, so our measurements characterize their global
star formation properties.

\subsection{O stars producing ionizing photons}
For those galaxies whose SEDs were fit with thermal components, we
used their fluxes at 33 GHz to calculate their thermal
luminosities. We then used those luminosities to calculate the number
of ionizing photons responsible for the thermal fluxes seen within the
GBT beam following Equation 2 in \citet{C92}:
\begin{equation}
\left(\frac{Q_{Lyc}}{s^{-1}}\right) \geq 6.3 \times 10^{52}
  \left(\frac{T_{e}}{10^{4}K}\right)^{-0.45} \left(\frac{\nu}{GHz}\right)^{0.1}
  \left(\frac{L_{T}}{10^{20}W Hz^{-1}}\right),
\end{equation}
where $Q_{Lyc}$ is the number of Lyman continuum photons emitted by
the region on thermal emission, $T_{e}$ is the electron temperature,
and $L_{T}$ is the thermal luminosity. The resulting values are
detailed in Table \ref{fig:sfrfractableunres} (unresolved galaxies)
and Table \ref{fig:sfrfractableres} (resolved galaxies). We used an
electron temperature of $10^{4}$K, as is typical for star-forming
regions \citep{C92}, and used $Q_{0} = 10^{49} s^{-1}$ as the number
of Lyman continuum photons emitted by an O7.5V star from Table 5 in
\citet{Vacca96}. We report the total number of O7.5V stars in the
galaxies that are unresolved by the GBT at 33 GHz, and the number of
O7.5V stars per square kiloparsec for the resolved galaxies in Tables
\ref{fig:sfrfractableunres} and \ref{fig:sfrfractableres}. As seen in
Table \ref{fig:sfrfractableunres}, the number of O7.5V stars in each
unresolved galaxy varies widely (log $\#$ O7.5V stars is between 2.42
and 4.66). This is likely due to the wide range in the unresolved
galaxies' overall star formation rates and physical areas observed.

\subsection{Supernova rates}
Since we were able to fit nonthermal components for all of our
galaxies, we calculated supernova rates ($\nu_{SN}$) for each of them
following Equation 18 in \citet{C92}:
\begin{equation}
\left(\frac{L_{N}}{10^{22}W Hz^{-1}}\right) \sim 13\left(\frac{\nu}{GHz}\right)^{-0.8} \left(\frac{\nu_{SN}}{yr^{-1}}\right),
\end{equation}
where $L_{N}$ is the nonthermal luminosity. We report the total
supernova rate of the unresolved galaxies in Table
\ref{fig:sfrfractableunres}, while for the resolved galaxies we
report the supernova rate per square kiloparsec in Table
\ref{fig:sfrfractableres}. The supernova rates of the unresolved
galaxies vary by three orders of magnitude (log SNe rate between -3.72
and -0.71), which is not surprising given the differences in star
formation rates and physical areas sampled.

\subsection{Star formation rates}
We calculated massive star formation rates (SFRs) from thermal fluxes
for each galaxy whose SEDs have a thermal component and from
nonthermal fluxes for all of our galaxies following Equations 21 and
23 of \citet{C92}:
\begin{equation}
\left(\frac{L_{N}}{W Hz^{-1}}\right) \sim 5.3 \times 10^{21}
  \left(\frac{\nu}{GHz}\right)^{-0.8} \left(\frac{SFR_{N}\left(M \geq 5M_{\sun}\right)}{M_{\sun} yr^{-1}}\right)
\end{equation}
\begin{equation}
\left(\frac{L_{T}}{W Hz^{-1}}\right) \sim 5.5 \times 10^{20}
  \left(\frac{\nu}{GHz}\right)^{-0.1} \left(\frac{SFR_{T}\left(M \geq 5M_{\sun}\right)}{M_{\sun}yr^{-1}}\right)
\end{equation}
where $L_{T}$ and $L_{N}$ are thermal and nonthermal luminosities,
respectively, calculated from each galaxy's thermal and nonthermal
fluxes, and $\nu = 33$ GHz. These equations are derived from Equations
2 and 18 of \citet{C92} (reproduced as Equations 2 and 3 in this
paper). Those equations were derived assuming (1) an extended
Miller-Scalo IMF \citep{MS79} with an exponent of $-2.5$, (2) that all
stars with masses greater than $8 \rm M_{\sun}$ become supernovae, and
(3) that dust absorption is negligible \citep{C92}. We then scaled the
massive SFRs generated by each equation by a factor of 5.6 to
transform them to total SFRs (M $\geq$ 0.1$M_{\sun}$) calculated with
a Kroupa IMF \citep{K01}. The galaxies' SFRs calculated from their
thermal and nonthermal fluxes are shown in Table
\ref{fig:sfrfractableunres} (unresolved galaxies) and Table
\ref{fig:sfrfractableres} (resolved galaxies). We report the total
massive SFRs of the unresolved galaxies, while we report the massive
SFR per square kiloparsec of the resolved galaxies. All of the
galaxies for which we calculated both thermal and nonthermal SFRs
showed agreement between the two to within an order of magnitude, but
not necessarily to within their margins of uncertainty. The
disagreement correlates with the thermal fractions of each galaxy:
galaxies with high thermal fractions were likely to have higher
thermal SFRs than nonthermal SFRs, while galaxies with low thermal
fractions showed the opposite relation. Like the differences in
thermal fractions between galaxies in our sample, disagreement could
be due to the different star formation timescales traced by the
thermal and nonthermal fluxes. Since these two emission components are
caused by physical processes that operate over differing lengths of
time (as discussed in Section 3.3.1), it is possible that the
discrepancies between the star formation rates could be used to infer
the recent star formation histories of the observed regions.

We compared the radio continuum SFRs to monochromatic SFRs from 24$\mu
m$ fluxes as described in \citet{Cal10}. The galaxies' SFRs (for the
unresolved galaxies) and SFR densities (for the resolved galaxies)
derived from 24$\mu m$ fluxes are listed in Table
\ref{fig:sfrfractableunres} and Table \ref{fig:sfrfractableres}. In
Figure~\ref{fig:comparesfrs}, we compare the SFRs derived from thermal
and nonthermal radio continuum fluxes of the unresolved galaxies for
which we fit two-component SEDs to SFRs derived from 24$\mu m$
fluxes. We find that most of the galaxies in our sample have higher
radio continuum SFRs (both from thermal and nonthermal fluxes) than
SFRs from 24$\mu m$ data. One possible explanation for this is that
extinction is lower at radio wavelengths than it is at 24$\mu
m$. Another possible explanation is that since radio continuum
emission traces very young star formation while 24$\mu m$ emission
traces less recent star formation, higher SFRs calculated from radio
continuum observations than from 24$\mu m$ data could be another
indication that our sample of galaxies is undergoing recent star
formation.

\subsection{Radio-far-infrared correlation}
There is a well-established tight correlation between far-infrared
(FIR) and radio flux in star-forming galaxies
\citep[e.g.][]{H85,M06,M12}. When plotted on a log-log scale, the
relationship between radio continuum and FIR flux for star-forming
galaxies appears linear. This correlation has been well-studied at low
frequencies ($\sim$ 1.4 GHz) where synchrotron emission is the
dominant component of radio emission in a star-forming galaxy. We
investigated whether this correlation could also be found on a global
scale at 33 GHz, where synchrotron emission is weaker than it is at
1.4 GHz and the relative contribution from thermal emission is more
significant.

We limited our study of the radio-FIR correlation to the galaxies in
our sample that are unresolved with the GBT beam at 33 GHz (as
discussed in Section 2.2). We chose this limit to ensure that we were
observing both the total area of radio emission and total area of
far-infrared emission in each galaxy. This minimizes issues related to
the different beam sizes of the GBT and IRAS (objects are considered
point sources to IRAS if they are more compact than 1$\arcmin$ at 60
$\rm \mu m$ and 2$\arcmin$ at 100 $\rm \mu m$). 

We fit a power law to our 33 GHz flux as a function of total FIR
flux. The total FIR flux was determined by a combination of archival
IRAS 100 $\rm \mu m$ and 60 $\rm \mu m$ fluxes as described in
\citet{H88} ($S_{FIR} = 2.58 S_{60\mu m} + S_{100 \mu m}$). We chose
to compute each galaxy's 33 GHz flux by taking the average of its
fluxes in the four sub-bands. We used this measure (rather than the
flux at 33 GHz inferred from the galaxies' SEDs) in order to eliminate
possible uncertainties in the flux due to using assumed spectral
indices in our fits. We found that the fluxes were related by $\rm log
\ S_{33} = (0.88 \pm 0.01) log \ S_{FIR} + log \ (5.3\times 10^{-4}
\pm 6\times 10^{-5})$. This correlation is relatively well-fit (the
fractional errors of both fit parameters are small) even though our
sample contains a wide range of thermal fractions. \citet{M12} found a
similar correlation between 33 GHz and $24 \rm \mu m$ fluxes for
resolved nuclei and individual star-forming regions of galaxies. We
find that the radio-FIR correlation at 33 GHz can be extended to
global measurements of galaxies' fluxes.

As a control of the tightness of the radio-FIR correlation in our
sample, we also fit a relationship between the galaxies' NVSS 1.4 GHz
fluxes and their total FIR fluxes. This relationship for the
unresolved galaxies in our sample is $\rm log \ S_{1.4} = (0.85 \pm
0.01) log \ S_{FIR} + log \ (0.0047 \pm 0.0006)$. The fractional
uncertainties on the fit parameters are similar to those of the fit at
33 GHz. We plot both correlations in Figure \ref{fig:rad-IR}.

As discussed in Section 3.2, we have determined thermal fractions from
SED fits assuming fixed thermal and nonthermal spectral indices. Due
to the limited number of radio data points we have for each galaxy, we
cannot more accurately constrain the thermal fractions at 33 GHz of
the galaxies in our sample at this time. Therefore, we do not have
enough information to definitively isolate thermal and nonthermal
components to explore whether the radio-FIR correlation is equally
tight for each. As an estimate, we have coded approximate thermal
fractions in the plot. Even given these limitations, we are confident
that a correlation exists between the unresolved galaxies' total radio
flux at 33 GHz and total FIR flux. \citet{M12} found a similar
correlation at 33 GHz for resolved nuclei and star-forming regions of
galaxies.

To further constrain the radio-FIR correlation at 33 GHz in our
sample, we calculated $q_{\nu}$ for each galaxy. $q_{\nu}$ is a
logarithmic measure of the ratio of total far-infrared flux ($S_{FIR}$
in Janskys) to radio continuum flux ($S_{\nu}$) in units of $\rm W
m^{-2} Hz^{-1}$ at a given frequency.  It is defined in \citet{H85} as
\begin{equation}
q_{\nu} = log\left(\frac{S_{FIR} \cdot 1.26 \times 10^{-14} W
  m^{-2}}{3.75 \times 10^{12} W m^{-2}}\right) -
log\left(\frac{S_{\nu}}{W m^{-2} Hz^{-1}}\right).
\end{equation}
The average $q_{33}$ for our sample is $q_{33}=3.3$, with a 1$\sigma$
scatter of 0.3. \citet{C92} reported that at 1.4 GHz, the average
value of $q_{1.4}$ from a large sample of galaxies is $q_{1.4}=2.3 \pm
0.2$. The average value of $q_{\nu}$ at 1.4 GHz for this set of
galaxies is $q_{1.4} = 2.4 \pm 0.2$, consistent with the \citet{C92}
value. Since $q_{\nu}$ is a function of the ratio of FIR flux to radio
flux at a given frequency, it makes sense that $q_{\nu}$ is larger
using 33 GHz fluxes than it is using 1.4 GHz fluxes (star-forming
galaxies are generally much brighter at 1.4 GHz than at 33 GHz). The
scatter on $q_{\nu}$ at 33 GHz is larger than that at 1.4 GHZ, which
indicates that the radio-FIR correlation is not as tight at 33 GHz as
at 1.4 GHz. This may be due to contamination from increased thermal
flux at 33 GHz. If the correlation is solely between synchrotron and
FIR emission, thermal flux at 33 GHz will increase the scatter in the
correlation. However, due to our small sample size, we cannot rule out
the possibility that the correlation is just as strong at 33 GHz,
where thermal fractions are higher, as it is at 1.4 GHz, where
nonthermal emission is typically much stronger. We note that the
galaxies with the highest thermal fractions lie above the fitted
correlation at 33 GHz, while the same is not true at 1.4 GHz, which
supports thermal emission being the cause of increased scatter.

In addition to plotting the radio-FIR correlation, we also plot the
ratio of 33 GHz flux to FIR flux, $q_{33}^{-1}$, against
$\alpha_{1.4-33}$ for our unresolved galaxies in Figure
\ref{fig:inverseqvsalpha}, similar to \citet{M12}. The plot shows an
increasing $q_{33}^{-1}$ for flatter values of
$\alpha_{1.4-33}$. Flatter $\alpha_{1.4-33}$ values are presumably
indicative of a higher proportion of thermal flux to nonthermal flux,
which is reflected in the highest thermal fractions in our sample also
having the flattest $\alpha_{1.4-33}$. A correlation between an
elevated $q_{33}^{-1}$ and flat values of $\alpha_{1.4-33}$ is not
surprising if the radio-FIR correlation is solely dependent on
synchrotron emission. If the radio-FIR correlation was independent of
the type of radio emission, $q_{33}^{-1}$ should be relatively
constant between galaxies and should not be affected by different
spectral indices or thermal fractions. Our data support that the
radio-FIR correlation is independent of a galaxy's thermal emission
since the addition of thermal emission results in elevated ratios of
33 GHz flux to FIR flux.

\subsubsection{Implications for star formation timescales}

When the timescales of the emission mechanisms for thermal,
nonthermal, and FIR fluxes are taken into account, the observed
relationship between the ratio of 33 GHz and FIR fluxes and
$\alpha_{1.4-33}$ may be a way to age-date an episode of star
formation. Since thermal flux is only produced by the shortest-lived
($\tau<10$ Myr) massive stars, its presence in large quantities
relative to synchrotron emission is indicative of very young star
formation. Since in addition to massive stars, infrared emission also
traces less massive stars ($\rm M > 5M_{\sun}$) that live longer than
the $\rm M > 8M_{\sun}$ stars that produce thermal and nonthermal
radio emission \citep{D90}, FIR emission is a tracer of star formation
on longer timescales. Stars with these masses can live up to $\sim$100
Myr, while nonthermal radio emission traces stars with lifetimes of up
to $\sim$30 Myr and whose emission is detectable for up to 100 Myr
\citep[for an illustrative plot of stellar lifetimes, see Figure 3
  of][]{R05}. In addition, infrared emission also contains a component
from diffuse dust that is heated by lower-mass stars with lifetimes
longer than 100 Myr. These timescales could mean that the galaxies
that show both flat spectral indices and enhanced $q_{33}^{-1}$ also
host the youngest areas of ongoing star formation. This correlation
could then be a method of determining approximate ages for galaxies'
global star formation. As a simple test, we used a Starburst 99 model
of a single instantaneous burst using default inputs (solar
metallicity, a 2-component Kroupa IMF, and no effects of cosmic ray
aging, escape, or absorption taken into account) run for 100 Myr
\citep{L99,V05,L10}. This model, depicted in Figure \ref{fig:sb99},
shows the flattest spectral indices and highest thermal fractions at
the earliest times of the starburst. Similarly, the steepest spectral
indices and lowest thermal fractions were seen as the lowest-mass
stars that produce supernovae were dying (at $\sim$40 Myr). The
starburst's ratio of 33 GHz luminosity to FIR luminosity was also high
at early times (between 3 Myr and 40 Myr) while the lowest ratios of
33 GHz luminosity to FIR luminosity were seen even later (after 40
Myr). While modeling a more robust quantitative relationship between
this observed correlation and the age of each galaxy's star-forming
episode is beyond the scope of this work (the simple model we used
does not take into account multiple co-existing generations of star
formation), the apparent relationship between enhanced 33 GHz flux,
flat spectral indices, and high thermal fractions is a promising
metric for future global radio and far-infrared photometric studies of
star-forming galaxies. Our simple model is not robust enough to
constrain the timescales' uncertainties, but is only meant to be
illustrative of a correlation visible in our data.

\section{Conclusions}

We have observed 27 local, well-studied, star-forming galaxies between
26-40 GHz with the GBT and obtained the first detections at this
frequency range for 22 of the galaxies. We determined the
contributions of thermal free-free and nonthermal synchrotron emission
to the galaxies' total radio emission. We have used these measures to
derive the number of massive, short-lived O stars and the number of
recent supernovae in the observed regions of each galaxy. In addition,
we have calculated SFRs for each galaxy using thermal and nonthermal
fluxes and explored the radio-FIR correlation for the unresolved
galaxies. We found that
\begin{itemize}
\item None of the galaxies have spectral incides indicative of purely
  thermal emission; eight galaxies show spectra that are too
    steep to fit thermal components,
\item Thermal fractions range from 10$\%$ to 90$\%$, with a
  median of $55\%$,
\item The radio-far infrared correlation holds for the unresolved
  galaxies at 1.4 GHz and 33 GHz, though the scatter at 33 GHz is
  larger due to the increased influence of thermal emission at higher
  frequencies, and
\item Galaxies with flat $\alpha_{1.4-33}$ and high thermal fractions
  have enhanced radio flux at 33 GHz with respect to far-infrared
  flux, which identifies them as galaxies with recent star
  formation. This is consistent with a simple model of a single
  starburst.
\end{itemize}

We found that the observed regions of our galaxies had a diverse mix
of radio continuum characteristics, with some galaxies' SEDs being
dominated at 33 GHz by the thermal emission indicative of ongoing
massive star formation, while others have little or no detectable
thermal emission. Even with this spread in the relative contributions
of thermal and nonthermal emission, we saw that there is still a
correlation between the global 33 GHz and far-infrared flux in the
unresolved galaxies. The scatter in the correlation is larger than
that at 1.4 GHz, likely due to the increased influence of thermal
emission at 33 GHz. We cannot, however, rule out that the radio-FIR
correlation is not solely dependent on synchrotron emission. We also
found that higher ratios of 33 GHz emission to FIR emission correlated
with flatter spectral indices (and higher thermal fractions) for
unresolved galaxies, which is consistent with younger ages in simple
starburst models. This correlation may be useful as a rough indicator
of the age of the most recent episode of star formation. Future global
studies of more homogeneous galaxy populations or resolved studies of
individual star-forming regions will enable better modeling of star
formation timescales using this metric.

In giving a broad measure of nearby galaxies' radio continuum
emission, our observations complement previous studies done with
interferometers in which individual star-forming regions in local
galaxies were highly resolved. With the GBT, we can simultaneously
observe compact and diffuse thermal and nonthermal emission and
determine their relative intensities, and in doing so estimate the
timescale for the current episode of star formation. Unfortunately, we
cannot make stricter timescale estimates than those discussed in
Section 3.3 at this time, as we do not have enough radio data points
to robustly fit thermal and nonthermal flux components with varying
spectral indices. Obtaining more unresolved radio fluxes at lower and
higher frequencies would help this effort.

This research has made use of the NASA/IPAC Extragalactic Database
(NED) which is operated by the Jet Propulsion Laboratory, California
Institute of Technology, under contract with the National Aeronautics
and Space Administration. We acknowledge the use of NASA's SkyView
facility (http://skyview.gsfc.nasa.gov) located at NASA Goddard Space
Flight Center. We thank the telescope operators and support staff at
the GBT for assistance with this project. K.R. acknowledges support
from an NRAO student observing support award (GSSP10-0002). K.R. also
thanks Brian Mason for his help with understanding the CCB observation
and data reduction process.

\begin{deluxetable}{l l l l l l l c l}
\rotate
\tablecolumns{9}
\tablewidth{0pt}
\tablecaption{Observation summary}
\tablehead{
\colhead{Source} &
\colhead{RA \tablenotemark{a}} &
\colhead{Dec} &
\colhead{D} &
\colhead{$R_{25}$ \tablenotemark{b}} &
\colhead{$R_{23\arcsec}$} &
\colhead{Hubble Type \tablenotemark{c}}&
\colhead{$\#$ nods \tablenotemark{d}}& 
\colhead{Ref code \tablenotemark{e}}\\
\colhead{} &
\colhead{(J2000)} &
\colhead{(J2000)} &
\colhead{(Mpc)} &
\colhead{(arcsec)} &
\colhead{(kpc)}&
\colhead{}&
\colhead{}&
\colhead{}\\
}
\startdata
NGC 520 & 01:24:34.8  &   +03:47:29 & 30.5 $\pm$ 2.1 & 122.2 $\pm$ 3.2 & 1.7 & Sa & 2 & B03\\
Maffei II & 02:41:55.9 &    +59:36:14 & 3.11 $\pm$ 0.23 & 20.8 $\pm$ 4.3 & 0.17 & SBbc & 2 & T94, T06\\
NGC 1222 & 03:08:56.8  &   -02:57:18 & 32.4 $\pm$ 2.3 & 47.5 $\pm$ 3.6 & 1.8 & E/S0 & 1 & B07\\
SBS 0335-052 & 03:37:44.0  &   -05:02:40 & 53.7 $\pm$ 3.8 & 13.8 \tablenotemark {e} & 3.0 & BCG/starburst \tablenotemark{g} & 2 & J09\\
IC 342 & 03:46:48.5 & +68:05:46 & 3.93 $\pm$ 0.27 & 599 $\pm$ 3.1 & 0.22 & SABc & 1 & M12, T06\\
NGC 1569 & 04:30:49.3  &   +64:50:52 & 1.39 $\pm$ 0.11 & 116.7 $\pm$ 3.2 & 0.08 & IB & 3 & L04\\
VII Zw 19 & 04:40:47.3  &   +67:44:09 & 67.4 $\pm$ 5.5 & 12.8 $\pm$ 3.6 & 3.8 &E & 3 & B00\\
NGC 1741 & 05:01:38.3  &   -04:15:24 & 54.6 $\pm$ 3.8 & 36.1 $\pm$ 3.6 & 3.0 & Sm & 1 & B00\\
II Zw 40 & 05:55:42.6  &   +03:23:32 & 11.1 $\pm$ 0.80 & 16.8 \tablenotemark{f} & 0.62 & Sbc & 3 & B02\\
Mrk 8 & 07:29:26.3  &   +72:07:44 & 52.5 $\pm$ 3.7 & 26.1 $\pm$ 4.1 & 2.9 & Sbc & 1 & B00\\
Mrk 86 & 08:13:14.6  &   +45:59:29 & 7.94 $\pm$ 1.5 & 61.3 $\pm$ 3.4 & 0.44 & SBm & 1 & A11\\
NGC 2903 & 09:32:09.7  &   +21:30:02 & 7.39 $\pm$ 0.52 & 360.8 $\pm$ 3.1 & 0.41 & SABb & 1 & T06\\
NGC 2997 & 09:45:38.7  &   -31:11:25 & 13.8 $\pm$ 0.90 & 307.0 $\pm$ 3.1 & 0.77 & SABc & 1 & K11\\
Mrk 1236 & 09:49:54.1  &   +00:36:58 & 27.6 $\pm$ 5.1 & 36.1 $\pm$ 3.3 & 1.5 & WR/HII \tablenotemark{g} & 1 & B00\\
NGC 3077 & 10:03:20.2  &   +68:44:01 & 2.55 $\pm$ 0.19 & 157.4 $\pm$ 3.2 & 0.14 & S? & 1 & M12, R05\\
NGC 3125 & 10:06:33.6  &   -29:56:08 & 13.8 $\pm$ 1.0 & 36.1 $\pm$ 3.3 & 0.77 & E & 1 & A11\\
Arp 233 & 10:32:31.1  &   +54:24:04 & 25.5 $\pm$ 1.8 & 34.4 $\pm$ 3.4 & 1.4 & I & 1 & T72, B00, A11\\
Arp 217 & 10:38:45.9  &   +53:30:11 & 19.2 $\pm$ 1.3 & 57.2 $\pm$ 3.3 & 1.1 & SABb & 1 & A11\\
Haro 3 & 10:45:22.4  &   +55:57:37 & 18.5 $\pm$ 1.3 & 40.5 $\pm$ 3.4 & 1.0 & Sb & 1 & T72, J04, A11\\
Pox 4 & 11:51:11.7  &   -20:35:57 & 52.5 $\pm$ 3.9 & 19.4 $\pm$ 3.4 & 2.9 & Sm & 1 & B00\\
NGC 4038 & 12:01:52.5  &   -18:52:02 & 21.5 $\pm$ 1.6 & 161.1 $\pm$ 3.2 & 1.2 & SBm & 1 & C04\\
NGC 4214 & 12:15:39.2  &   +36:19:41 & 2.94 $\pm$ 0.27 & 202.8 $\pm$ 3.2 & 0.16 & IB & 1 & B00\\
NGC 4449 & 12:28:10.1  &   +44:05:33 & 3.54 $\pm$ 0.25 & 140.3 $\pm$ 3.2 & 0.20 & IB & 1 & R08\\
NGC 4490 & 12:30:36.7  &   +41:38:26 & 9.22 $\pm$ 0.65 & 202.8 $\pm$ 3.3 & 0.51 & SBcd & 1 & A11\\
Tol 35 & 13:27:07.2  &   -27:57:26 & 25.2 $\pm$ 1.8 & 43.4 $\pm$ 3.4 & 1.4 & Sab & 1 & B00\\
M 51 & 13:29:52.4  &   +47:11:40 & 7.90 $\pm$ 0.85 & 414.1 $\pm$ 3.1 & 0.44 & Sbc & 1 & M12, T94, D11\\
M 101 & 14:03:12.5  &   +54:20:53 & 6.46 $\pm$ 0.18 & 719.7 $\pm$ 3.1 & 0.36 & SABc & 1 & M12, T94\\
\enddata
\tablenotetext{a}{RA and Dec are center positions of the beam for each observation.}
\tablenotetext{b}{$R_{25}$ values are derived from average $d_{25}$ from Hyperleda (http://leda.univ-lyon1.fr/).}
\tablenotetext{c}{Hubble types from Hyperleda except where specified.}
\tablenotetext{d}{Each nod is 70 seconds long.}
\tablenotetext{e}{These galaxies have previously published radio continuum observations. M12 denotes galaxies detected with the CCB by \citet{M12} at 33 GHz. T72 denotes galaxies that were observed, but not detected, by \citet{T72} at 9.5 mm (31.6 GHz). T94 \citep[]{TH94}, B00 \citep[]{B00}, B02 \citep[]{B02}, B03 \citep[]{B03}, C04 \citep[]{C04}, J04 \citep[]{J04}, L04 \citep[]{L04}, R05 \citep[]{RG05}, T06 \citep[]{T06}, B07 \citep[]{B07}, R08 \citep[]{R08a}, J09 \citep[]{J09}, D11 \citep[]{D11}, K11 \citep[]{Kod11}, and A11 \citep[]{A11} denote galaxies that were observed at lower frequencies. The galaxies are described in more detail in these papers.}
\tablenotetext{f}{No radius data available from Hyperleda so major axis diameter  from NED is reported.}
\tablenotetext{g}{No Hubble type data available from Hyperleda so classification listed on NED is reported.}
\label{fig:obssummary}
\end{deluxetable}

\begin{deluxetable}{l l l l l}
\tablecolumns{5}
\tablewidth{0pt}
\tablecaption{Observed Flux}
\tablehead{
\colhead{Source} &
\colhead{27.75 GHz} &
\colhead{31.25 GHz} &
\colhead{34.75 GHz} &
\colhead{38.25 GHz} \\
\colhead{} &
\colhead{(mJy)} &
\colhead{(mJy)} &
\colhead{(mJy)} &
\colhead{(mJy)} \\
}
\startdata 
\sidehead{Four-sub-band detections: flux in each sub-band}
NGC 520 & $21.77 \pm 0.39$ & $19.50 \pm 0.19$ & $17.44 \pm 0.27$ & $15.43 \pm 0.43$ \\ 
Maffei II & $23.17 \pm 0.28$ & $19.42 \pm 0.16$ & $16.36 \pm 0.20$ & $13.98 \pm
0.29$\\ 
NGC 1222 & $9.83 \pm 0.53$ & $8.57 \pm 0.25$ & $7.91 \pm 0.36$ & $7.26 \pm 0.54$ \\ 
SBS 0335-052 & $0.66 \pm 0.20$ & $0.62 \pm 0.12$ & $0.69 \pm 0.18$ & $0.52
\pm 0.24$ \\ 
IC 342 & $35.59 \pm 3.46$ & $31.03 \pm 0.87$ & $27.82 \pm 1.08$ & $25.25 \pm 2.21$ \\ 
NGC 1569 & $28.60 \pm 0.19$ & $24.67 \pm 0.12$ & $21.31 \pm 0.15$ & $18.75 \pm 0.21$ \\ 
VII Zw 19 & $3.16 \pm 0.27$ & $2.60 \pm 0.14$ & $2.26 \pm 0.19$ & $2.11 \pm 0.27$ \\ 
NGC 1741 & $2.12 \pm 0.31$ & $2.07 \pm 0.18$ & $1.67 \pm 0.22$ & $0.84 \pm 0.32$ \\ 
II Zw 40 & $15.09 \pm 2.00$ & $14.12 \pm 0.50$ & $12.99 \pm 0.62$ & $12.20 \pm 1.28$ \\ 
Mrk 8 & $3.23 \pm 0.38$ & $2.84 \pm 0.39$ & $2.21 \pm 0.33$ & $2.53 \pm 0.42$\\
NGC 2903 & $14.50 \pm 1.47$ & $12.80 \pm 0.52$ & $11.15 \pm 0.82$ & $9.60 \pm 1.16$ \\ 
NGC 2997 & $5.14 \pm 1.12$ & $4.77 \pm 0.45$ & $4.38 \pm 0.63$ & $3.90 \pm 0.85$ \\ 
NGC 3077 & $6.94 \pm 0.39$ & $5.93 \pm 0.38$ & $5.45 \pm 0.32$ & $4.71 \pm 0.43$ \\ 
NGC 3125 & $8.15 \pm 1.03$ & $7.43 \pm 0.44$ & $6.52 \pm 0.58$ & $5.92 \pm 0.79$ \\ 
Arp 233 & $4.22 \pm 0.43$ & $4.05 \pm 0.40$ & $3.49 \pm 0.31$ & $2.93 \pm 0.43$ \\ 
Arp 217 & $25.98 \pm 0.43$ & $21.65 \pm 0.40$ & $19.04 \pm 0.31$ & $16.26 \pm 0.43$ \\ 
Haro 3 & $6.10 \pm 0.43$ & $5.49 \pm 0.40$ & $4.78 \pm 0.31$ & $4.35 \pm 0.43$ \\ 
NGC 4038 & $7.77 \pm 0.96$ & $5.44 \pm 0.42$ & $4.64 \pm 0.55$ & $4.27 \pm 0.74$ \\ 
NGC 4214 & $7.18 \pm 0.93$ & $6.22 \pm 0.38$ & $5.26 \pm 0.48$ & $4.68 \pm 0.69$ \\ 
NGC 4449 & $4.37 \pm 0.90$ & $3.45 \pm 0.38$ & $2.73 \pm 0.44$ & $2.30 \pm 0.65$ \\ 
NGC 4490 & $3.40 \pm 0.90$ & $2.64 \pm 0.38$ & $1.64 \pm 0.46$ & $1.33 \pm 0.66$ \\ 
M 51 & $7.67 \pm 0.94$ & $5.84 \pm 0.38$ & $4.89 \pm 0.47$ & $3.65 \pm 0.65$ \\ 
\hline 
\sidehead{Marginal  detections: average of four sub-bands' fluxes}
Mrk 86 & 0.42 $\pm$ 0.19 & & & \\ 
Mrk 1236 & 0.99 $\pm$ 0.40 & & & \\ 
Pox 4 & 1.62 $\pm$ 0.29 & & & \\ 
Tol 35 & $< 0.87$ & & & \\ 
M 101\tablenotemark{b} & 0.69 $\pm$ 0.27 & & & \\ 
\enddata 

\tablenotetext{a}{$\alpha_{26-40}$ is calculated using the
  27.75 GHz and 38.25 GHz fluxes.}

\tablenotetext{b}{The lower-frequency radio continuum emission of M
  101 is more extended than the separation between the on-source and
  off-source beams \citep{G90}, so its reported flux may suffer from
  oversubtraction due to emission in the off-source beam.}
\label{fig:uncorrectedflux}

\end{deluxetable}

\begin{deluxetable}{l l l l l}
\tablecolumns{5}
\tablewidth{0pt}
\tablecaption{Beam correction factors for galaxies detected in all four sub-bands}
\tablehead{
\colhead{Source} &
\colhead{27.75 GHz} &
\colhead{31.25 GHz} &
\colhead{34.75 GHz} &
\colhead{38.25 GHz} \\
\colhead{Beamsize} &
\colhead{26.7 $\arcsec$} &
\colhead{23.7 $\arcsec$} &
\colhead{21.3 $\arcsec$} &
\colhead{19.4 $\arcsec$} \\
}

\startdata
NGC 520 & 0.99 & 0.99 & 1 & 1.01\\
Maffei II & 0.89 & 0.95 & 1 & 1.05\\
NGC 1222\tablenotemark{a} & \nodata & \nodata & \nodata & \nodata\\
SBS 0335-052 & 0.98 & 0.99 & 1 & 1.01\\
IC 342 & 0.93 & 0.97 & 1 & 1.03\\
NGC 1569 & 0.81 & 0.91 & 1 & 1.09\\
VII Zw 19 & 0.97 & 0.99 & 1 & 1.02\\
NGC 1741 & 0.98 & 0.99 & 1 & 1.01\\
II Zw 40 & 0.95 & 0.97 & 1 & 1.03\\
Mrk 8 & 0.92 & 0.96 & 1 & 1.04\\
NGC 2903 & 0.96 & 0.98 & 1 & 1.03\\
NGC 2997 & 0.89 & 0.94 & 1 & 1.07\\
NGC 3077 & 0.97 & 0.98 & 1 & 1.02\\
NGC 3125 & 0.92 & 0.96 & 1 & 1.04\\
Arp 233 & 0.98 & 0.99 & 1 & 1.01\\
Arp 217 & 0.87 & 0.93 & 1 & 1.07\\
Haro 3 & 0.95 & 0.98 & 1 & 1.03\\
NGC 4038 & 0.83 & 0.94 & 1 & 1.03\\
NGC 4214 & 0.87 & 0.94 & 1 & 1.07\\
NGC 4449 & 0.78 & 0.89 & 1 & 1.11\\
NGC 4490 & 0.75 & 0.87 & 1 & 1.13\\
M 51 & 0.83 & 0.91 & 1 & 1.08\\

\hline
average & 0.91 & 0.95 & 1 & 1.05\\
standard dev & 0.07 & 0.04 & 0 & 0.04\\
\enddata
\tablenotetext{a}{NGC 1222 did not have archival radio data available for re-imaging, so average beam correction values were used.}
\label{fig:beamcorrections}
\end{deluxetable}

\begin{deluxetable}{l l l l l l l}
\tablecolumns{7}
\tablewidth{0pt}
\tablecaption{Corrected Flux}
\tablehead{
\colhead{Source} &
\colhead{27.75 GHz} &
\colhead{31.25 GHz} &
\colhead{34.75 GHz} &
\colhead{38.25 GHz} &
\colhead{$\alpha_{26-40}$\tablenotemark{a}} &
\colhead{Notes\tablenotemark{b}}\\
\colhead{} &
\colhead{(mJy)} &
\colhead{(mJy)} &
\colhead{(mJy)} &
\colhead{(mJy)} &
\colhead{} &
\colhead{} \\
}
\startdata
\sidehead{Four-port detections: flux in each port}
NGC 520 & 21.54 $\pm$ 0.39 & 19.30 $\pm$ 0.19 & 17.44 $\pm$ 0.27 & 15.58 $\pm$ 0.43 & -1.01 $\pm$ 0.10 & \\
Maffei II & 20.62 $\pm$ 0.28 & 18.45 $\pm$ 0.16 & 16.36 $\pm$ 0.20 & 14.68 $\pm$ 0.29 & -1.06 $\pm$ 0.07 & 1\\
NGC 1222 & 8.94 $\pm$ 0.53 & 8.14 $\pm$ 0.25 & 7.91 $\pm$ 0.36 & 7.62 $\pm$ 0.54 & -0.50 $\pm$ 0.29 & 2\\
SBS 0335-052 & 0.65 $\pm$ 0.20 & 0.61 $\pm$ 0.12 & 0.69 $\pm$ 0.18 & 0.53 $\pm$ 0.24 & -0.65 $\pm$ 1.7\\
IC 342 & 33.10 $\pm$ 3.5 & 30.10 $\pm$ 0.87 & 27.82 $\pm$ 1.1 & 26.01 $\pm$ 2.2 & -0.75 $\pm$ 0.42 & \\
NGC 1569 & 23.17 $\pm$ 0.19 & 22.45 $\pm$ 0.12 & 21.31 $\pm$ 0.15 & 20.44 $\pm$ 0.21 & -0.39 $\pm$ 0.04 & 1\\
VII Zw 19 & 3.07 $\pm$ 0.27 & 2.54 $\pm$ 0.14 & 2.26 $\pm$ 0.19 & 2.15 $\pm$ 0.27 & -1.10 $\pm$ 0.48\\
NGC 1741 & 2.08 $\pm$ 0.31 & 2.05 $\pm$ 0.18 & 1.67 $\pm$ 0.22 & 0.85 $\pm$ 0.32 & -2.78\tablenotemark{2} $\pm$ 1.3 & \\
II Zw 40 & 14.33 $\pm$ 2.0 & 13.70 $\pm$ 0.50 & 12.99 $\pm$ 0.62 & 12.59 $\pm$ 1.3 & -0.41 $\pm$ 0.54\\
Mrk 8 & 2.97 $\pm$ 0.38 & 2.72 $\pm$ 0.39 & 2.21$\pm$ 0.33 & 2.63$\pm$ 0.42 & -0.38 $\pm$ 0.64 & \\
NGC 2903 & 13.91 $\pm$ 1.5 & 12.54 $\pm$ 0.52 & 11.15 $\pm$ 0.82 & 9.89 $\pm$ 1.2 & -1.06 $\pm$ 0.51 & \\
NGC 2997 & 4.58 $\pm$ 1.1 & 4.48 $\pm$ 0.45 & 4.38 $\pm$ 0.63 & 4.17 $\pm$ 0.85 & -0.29 $\pm$ 0.98 & 1\\
NGC 3077 & 6.73 $\pm$ 0.39 & 5.81 $\pm$ 0.38 & 5.45 $\pm$ 0.32 & 4.80 $\pm$ 0.43 & -1.05 $\pm$ 0.33 & \\
NGC 3125 & 7.49 $\pm$ 1.0 & 7.13 $\pm$ 0.44 & 6.52 $\pm$ 0.58 & 6.15 $\pm$ 0.79 & -0.62 $\pm$ 0.58 & \\
Arp 233 & 4.14 $\pm$ 0.43 & 4.01 $\pm$ 0.40 & 3.49 $\pm$ 0.31 & 2.96 $\pm$ 0.43 & -1.05 $\pm$ 0.56 & \\
Arp 217 & 22.60 $\pm$ 0.43 & 20.14 $\pm$ 0.40 & 19.04 $\pm$ 0.31 & 17.40 $\pm$ 0.43 & -0.81 $\pm$ 0.10 & \\
Haro 3 & 5.79 $\pm$ 0.43 & 5.38 $\pm$ 0.40 & 4.78 $\pm$ 0.31 & 4.48 $\pm$ 0.43 & -0.80 $\pm$ 0.38\\
NGC 4038 & 6.45 $\pm$ 0.96 & 5.11 $\pm$ 0.42 & 4.64 $\pm$ 0.55 & 4.39 $\pm$ 0.74 & -1.20 $\pm$ 0.70 & 1\\
NGC 4214 & 6.25 $\pm$ 0.93 & 5.84 $\pm$ 0.38 & 5.26 $\pm$ 0.48 & 5.01 $\pm$ 0.69 & -0.69 $\pm$ 0.63 & 1\\
NGC 4449 & 3.41 $\pm$ 0.90 & 3.07 $\pm$ 0.38 & 2.73 $\pm$ 0.44 & 2.56 $\pm$ 0.65 & -0.90 $\pm$ 1.1 & 1\\
NGC 4490 & 2.55 $\pm$ 0.90 & 2.30 $\pm$ 0.38 & 1.64 $\pm$ 0.46 & 1.50 $\pm$ 0.66 & -1.65 $\pm$ 1.8 & 1\\
M 51 & 6.36 $\pm$ 0.94 & 5.32 $\pm$ 0.38 & 4.89 $\pm$ 0.47 & 3.94 $\pm$ 0.65 & -1.49 $\pm$ 0.69 & 1,3\\
\enddata 
\tablenotetext{a}{$\alpha_{26-40}$ is calculated using the
  27.75 GHz and 38.25 GHz fluxes.}
  
\tablenotetext{b}{Notes: 1: Lower-frequency radio continuum emission
  is more extended than the largest beam. 2: No lower-frequency radio
  continuum emission available; used characteristic beam-size
  correction values. 3: The lower-frequency radio continuum emission
  of M 51 is more extended than the separation between the on-source
  and off-source beams \citep{K84}, so its reported flux may suffer
  from oversubtraction due to emission in the off-source beam.}

\tablenotetext{c}{Though $\alpha_{26-40}$ for NGC 1741 is extremely
  steep (determining $\alpha_{26-40}$ via calculation and via a fit
  both result in $\alpha_{26-40} < -2$), its SED can be fit with a
  spectral index of $\alpha = -0.8$ when a 1.4 GHz data point is
  incorporated into the fit.}

\label{fig:correctedflux}

\end{deluxetable}

\begin{deluxetable}{l l l l l l l}
\small     
\rotate     
\tablecolumns{7}     
\tablewidth{0pt}     
\tablecaption{Star formation properties of unresolved galaxies}     
\tablehead{     
\colhead{Source} &     
\colhead{Thermal fraction \tablenotemark{a}} &     
\colhead{Log max \#} &     
\colhead{Log SNe rate} &
\colhead{Thermal SFR \tablenotemark{b}} &
\colhead{Nonthermal SFR \tablenotemark{c}} &
\colhead{Infrared SFR \tablenotemark{d}}\\     
\colhead{} &     
\colhead{} &     
\colhead{O7.5V stars} &     
\colhead{($yr^{-1}$)} &
\colhead{($M_{\sun} yr^{-1}$)} & 
\colhead{($M_{\sun} yr^{-1}$)} &
\colhead{($M_{\sun} yr^{-1}$)}\\     
}     
\startdata     
NGC 520 &  0.24 $\pm$ 0.01 &  4.65 &  -0.71 &  7.16 $\pm$ 0.43 &  26.7 $\pm$ 3.7 & 5.99 $\pm$ 0.94\\
NGC 1222 &  0.43 $\pm$ 0.02 &  4.59 &  -1.14 &  6.26 $\pm$ 0.48 &  9.8 $\pm$ 1.4 & 5.03 $\pm$ 0.79\\
SBS0335-052 &  0.79 $\pm$ 0.06 &  4.18 &  -2.25 &  2.43 $\pm$ 0.50 &  0.77 $\pm$ 0.25 & \nodata\\
IC 342 &  0.53 $\pm$ 0.02 &  3.41 &  -2.50 &  0.41 $\pm$ 0.03 &  0.43 $\pm$ 0.06 & 0.60 $\pm$ 0.09\\
VII Zw 19 &  0.38 $\pm$ 0.003 &  4.66 &  -0.97 &  7.28 $\pm$ 0.07 &  14.4 $\pm$ 2.3 & 16.6 $\pm$ 3.0\\
NGC 1741 &  $<$ 1.00 &  $<$ 4.87 &  -1.03 &  $<$ 12.0 &  12.7 $\pm$ 1.8 & 3.80 $\pm$ 0.62\\
II Zw 40 &  0.91 $\pm$ 0.01 &  4.20 &  -2.63 &  2.57 $\pm$ 0.09 &  0.32 $\pm$ 0.05 & 0.46 $\pm$ 0.08\\
Mrk 8 &  0.50 $\pm$ 0.05 &  4.58 &  -1.26 &  6.1 $\pm$ 1.1 &  7.4 $\pm$ 1.2 & 2.19 $\pm$ 0.38\\
NGC 2903 &  $<$ 0.94 &  $<$ 3.90 &  -1.92 &  $<$ 1.28 &  1.62 $\pm$ 0.23 & 0.31 $\pm$ 0.05\\
NGC 3077 &  0.66 $\pm$ 0.02 &  2.42 &  -3.72 &  0.042 $\pm$ 0.003 &  0.026 $\pm$ 0.004 & 0.0022 $\pm$ 0.004\\
NGC 3125 &  0.77 $\pm$ 0.02 &  4.03 &  -2.35 &  1.73 $\pm$ 0.11 &  0.61 $\pm$ 0.10 & 0.30 $\pm$ 0.06\\
Arp 233 &  0.71 $\pm$ 0.02 &  4.25 &  -1.99 &  2.89 $\pm$ 0.22 &  1.40 $\pm$ 0.21 & 1.29 $\pm$ 0.21\\
Arp 217 &  $<$ 0.99 &  $<$ 4.92 &  -0.92 &  $<$ 13.4 &  16.2 $\pm$ 2.1 & 3.71 $\pm$ 0.61\\
Haro 3 &  0.85 $\pm$ 0.02 &  4.19 &  -2.40 &  2.51 $\pm$ 0.13 &  0.54 $\pm$ 0.10 & 0.65 $\pm$ 0.11\\

\hline
\sidehead{Marginal Detections}
Mrk 86 &  \nodata &  $<$ 2.82 &  $<$ -3.02 &  $<$ 0.11 &  $<$ 0.13 & 0.037 $\pm$ 0.015\\
Mrk 1236 &  \nodata &  $<$ 4.25 &  $<$ -1.60 &  $<$ 2.88 &  $<$ 3.43 & 0.36 $\pm$ 0.15\\
Pox 4 &  \nodata &  $<$ 4.87 &  $<$ -0.98 &  $<$ 11.8 &  $<$ 14.1 & 0.87 $\pm$ 0.27\\
Tol 35 &  \nodata &  $<$ 3.99 &  $<$ -1.86 &  $<$ 1.57 &  $<$ 1.87 & \nodata\\

\enddata     
     
\tablenotetext{a}{The thermal fractions of each galaxy at 33 GHz are
  based on two-component fits, except for the galaxies with negative
  fitted thermal components. In such cases, the thermal fractions are
  presented as $3\sigma$ upper limits.}

\tablenotetext{b}{Thermal SFRs are calculated using thermal flux following Equation 4 scaled to a Kroupa IMF.} 
 
\tablenotetext{c}{Nonthermal SFRs are calculated using nonthermal flux following Equation 5 scaled to a Kroupa IMF.}

\tablenotetext{d}{Infrared SFRs are calculated from IRAS 24$\mu m$ fluxes using Equations 1 and 17 of \citet{Cal10}.}

\label{fig:sfrfractableunres}     
     
\end{deluxetable}     

\begin{deluxetable}{l l l l l l l}
\small
\rotate
\tablecolumns{7}
\tablewidth{0pt}
\tablecaption{Star formation properties of resolved galaxies}
\tablehead{
\colhead{Source} &
\colhead{Thermal fraction \tablenotemark{a}} &
\colhead{Log max \#} &
\colhead{Log SNe rate} &
\colhead{Thermal SFR} &
\colhead{Nonthermal SFR} &
\colhead{Infrared SFR}\\
\colhead{} &
\colhead{}&
\colhead{O7.5V stars }&
\colhead{}&
\colhead{density \tablenotemark{b}}&
\colhead{density \tablenotemark{c}}&
\colhead{density \tablenotemark{d}}\\
\colhead{} &
\colhead{} &
\colhead{($kpc^{-2}$)} &
\colhead{($yr^{-1} kpc^{-2}$)} &
\colhead{($M_{\sun} yr^{-1} kpc^{-2}$)} &
\colhead{($M_{\sun} yr^{-1} kpc^{-2}$)} &
\colhead{($M_{\sun} yr^{-1} kpc^{-2}$)}\\
}
\startdata
MaffeiII &  $<$ 0.97 &  $<$ 4.27 &  -1.57 &  $<$ 2.99 &  3.69 $\pm$ 0.77 & 0.039 $\pm$ 0.009\\
NGC 1569 &  0.11 $\pm$ 0.03 &  3.41 &  -1.52 &  0.41 $\pm$ 0.15 &  4.06 $\pm$ 0.92 & 0.31 $\pm$ 0.07\\
NGC 2997 &  0.32 $\pm$ 0.01 &  3.19 &  -2.34 &  0.25 $\pm$ 0.04 &  0.63 $\pm$ 0.12 & \nodata\\
NGC 4038 &  $<$ 0.87 &  $<$ 3.84 &  -1.95 &  $<$ 1.11 &  1.53 $\pm$ 0.33 & \nodata\\
NGC 4214 &  0.58 $\pm$ 0.03 &  3.55 &  -2.44 &  0.56 $\pm$ 0.12 &  0.49 $\pm$ 0.13 & 0.076 $\pm$ 0.022\\
NGC 4449 &  $<$ 1.00 &  $<$ 3.65 &  -2.25 &  $<$ 0.72 &  0.76 $\pm$ 0.15 & \nodata\\
NGC 4490 &  $<$ 0.92 &  $<$ 3.52 &  -2.29 &  $<$ 0.53 &  0.69 $\pm$ 0.15 & 0.16 $\pm$ 0.04\\
M 51 &  $<$ 0.94 &  $<$ 3.84 &  -1.98 &  $<$ 1.11 &  1.41 $\pm$ 0.43 & 0.13 $\pm$ 0.04\\

\hline
\sidehead{Marginal Detection}
M 101 &  \nodata &  $<$ 2.83 &  $<$ -3.02 &  $<$ 0.26 &  $<$ 0.32 & 0.015 $\pm$ 0.002\\
\enddata

\tablenotetext{a}{The thermal fractions of each galaxy at 33 GHz are
  based on two-component fits, except for the galaxies with negative
  fitted thermal components. In such cases, the thermal fractions
  are presented as $3\sigma$ upper limits.}

\tablenotetext{b}{Thermal SFR densities are calculated
  using thermal flux following Equation 4 and scaled to a Kroupa IMF over the area of the 23$\arcsec$ GBT beam.}

\tablenotetext{c}{Nonthermal SFR densities are calculated
  using nonthermal flux following Equation 5 and scaled to a Kroupa IMF over the area of the 23$\arcsec$ GBT beam.}

\tablenotetext{d}{Infrared SFR densities are calculated from IRAS 24$\mu m$ fluxes using Equations 1 and 17 of \citet{Cal10} over the area of the 47$\arcsec$ IRAS beam at 25$\mu m$.}

\label{fig:sfrfractableres}

\end{deluxetable}

\begin{deluxetable}{l l l l l l}     
\tablecolumns{6}     
\tablewidth{0pt}     
\tablecaption{Radio and far-infrared properties of unresolved galaxies}     
\tablehead{     
\colhead{Source} &     
\colhead{33 GHz flux\tablenotemark{a}} &     
\colhead{1.4 GHz flux\tablenotemark{b}} &     
\colhead{Far-IR flux\tablenotemark{c}} &
\colhead{$\alpha_{1.4-33}$} &
\colhead{$q_{33}$\tablenotemark{d}} \\     
\colhead{} &     
\colhead{(mJy)} &     
\colhead{(mJy)} &     
\colhead{(Jy)} &
\colhead{} & 
\colhead{} \\     
}     
\startdata
NGC 520 & $18.5 \pm 0.2$ & $176.3 \pm 5.3$ & $129.72 \pm 0.04$ & $-0.71 \pm 0.01$ & $3.37 \pm 0.004$\\
NGC 1222 & $8.15 \pm 0.22$ & $61.7 \pm 1.9$ & $47.89 \pm 0.01$ & $-0.64 \pm 0.01$ & $3.30 \pm 0.01$\\
SBS 0335-052 & $0.62 \pm 0.04$ & $2.3 \pm 0.4$ & \nodata & $-0.41 \pm 0.06$ & \nodata\\
IC 342 & $29.3 \pm 1.1$ & $190.7 \pm 7.3$ & $348.22 \pm 0.03$ & $-0.59 \pm 0.02$ & $3.60 \pm 0.02$\\
VII Zw 19 & $2.51 \pm 0.11$ & $20.5 \pm 0.7$ & $18.97 \pm 0.03$ & $-0.66 \pm 0.02$ & $3.40 \pm 0.02$\\
NGC 1741 & $1.66 \pm 0.13$ & $30.4 \pm 1.6$ & $16.13 \pm 0.02$ & $-0.92 \pm 0.03$ & $3.51 \pm 0.03$\\
II Zw 40 & $13.4 \pm 0.6$ & $32.5 \pm 1.1$ & $22.87 \pm 0.03$ & $-0.28 \pm 0.02$ & $2.76 \pm 0.02$\\
Mrk 8 & $2.63 \pm 0.19$ & $18.1 \pm 1.0$ & $348.22 \pm 0.03$ & $-0.61 \pm 0.03$ & $3.11 \pm 0.03$\\
Mrk 86 & $0.42 \pm 0.19$ & $10.5 \pm 1.7$ & $14.75 \pm 0.03$ & $-1.02 \pm 0.15$ & $4.07 \pm 0.20$\\
NGC 2903 & $11.9 \pm 0.5$ & $444.5 \pm 13.9$ & $176.93 \pm 0.03$ & $-1.14 \pm 0.02$ & $3.70 \pm 0.02$\\
Mrk 1236 & $0.99 \pm 0.40$ & $16.5 \pm 1.8$ & $11.03 \pm 0.03$ & $-0.89 \pm 0.13$ & $3.57 \pm 0.18$\\
NGC 3077 & $5.70 \pm 0.19$ & $29.0 \pm 1.5$ & $61.65 \pm 0.02$ & $-0.51 \pm 0.02$ & $3.56 \pm 0.01$\\
NGC 3125 & $6.82 \pm 0.37$ & $26.6 \pm 1.6$ & $19.31 \pm 0.03$ & $-0.43 \pm 0.03$ & $2.98 \pm 0.02$ \\
Arp 233 & $3.65 \pm 0.20$ & $16.6 \pm 0.6$ & $17.84 \pm 0.03$ & $-0.48 \pm 0.02$ & $3.22 \pm 0.02$\\
Arp 217 & $19.8 \pm 0.2$ & $362.8 \pm 12.4$ & $127.65 \pm 0.03$ & $-0.92 \pm 0.01$ & $3.34 \pm 0.004$\\
Haro 3 & $5.11 \pm 0.20$ & $15.5 \pm 0.9$ & $19.51 \pm 0.02$ & $-0.35 \pm 0.02$ & $3.11 \pm 0.02$ \\
Pox 4 & $1.62 \pm 0.29$ & $4.2 \pm 0.5$ & \nodata & $-0.30 \pm 0.07$ & \nodata\\
\enddata 

\tablenotetext{a}{\small The 33 GHz flux is the average of the
  corrected fluxes in the four sub-bands.}
\tablenotetext{b}{\small The 1.4 GHz flux is taken from the NRAO VLA
  Sky Survey.}
\tablenotetext{c}{\small The far-IR flux is derived from IRAS $60 \mu
  m$ and $100 \mu m$ fluxes using $S_{FIR} = 2.58 S_{60 \mu m} +
  S_{100 \mu m}$ \citep{H88}.}
\tablenotetext{d}{\small $q_{33}$ is a logarithmic measure of the
  ratio of far-IR flux (in Jy) to 33 GHz flux (in $\rm W m^{-2}
  Hz^{-1}$).}

\label{fig:radioFIRprops}
\end{deluxetable}

\begin{figure}
\begin{center}
\includegraphics[width=0.75\linewidth]{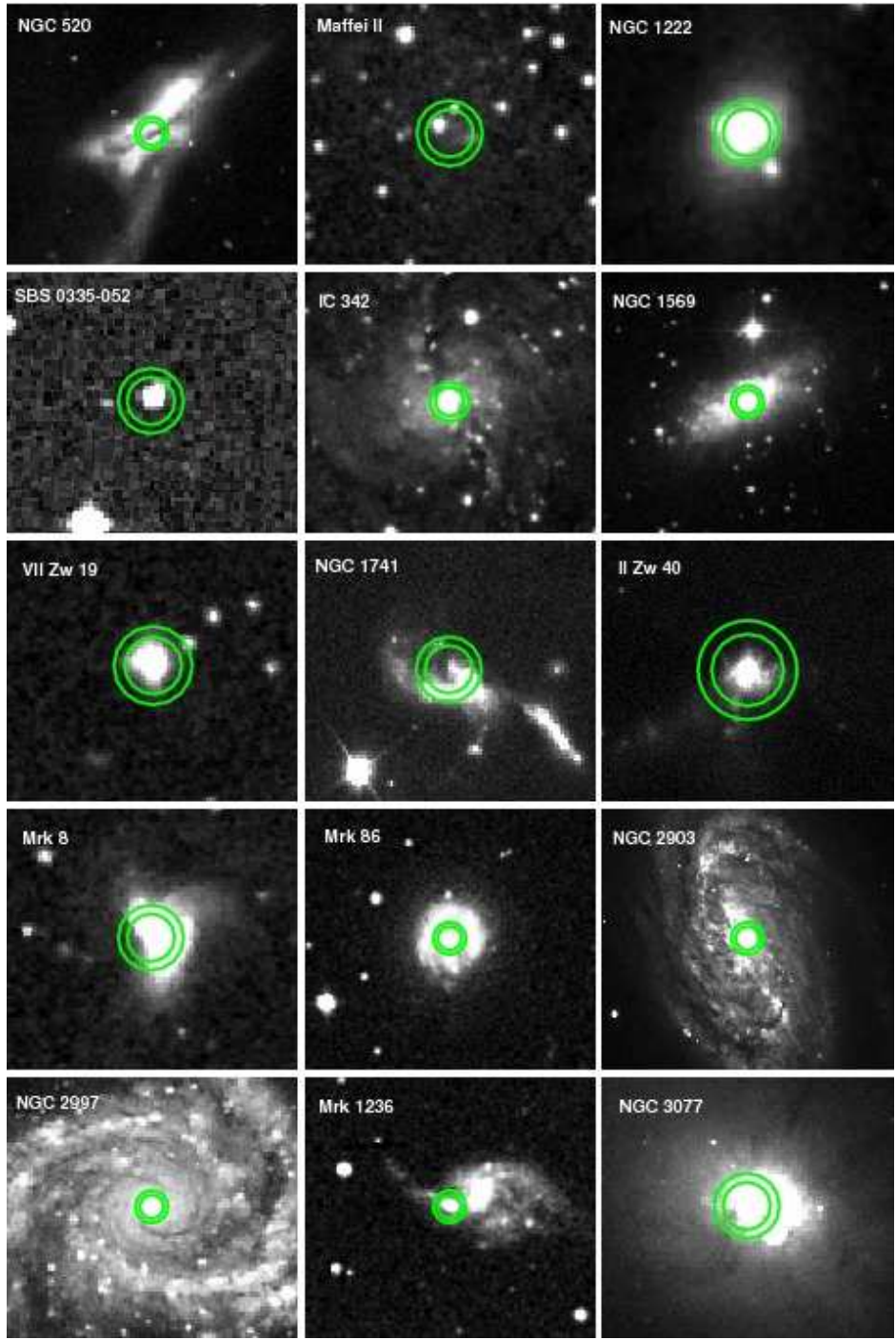}
\end{center}
\caption{Largest ($\sim 27\arcsec$) and smallest ($\sim 19\arcsec$)
  beam sizes overlaid on SDSS g or DSS B images of each galaxy
  observed. The galaxies are presented in the order listed in Table 1
  viewed left to right and top to bottom.}
\label{fig:beamsizes}
\end{figure}

\begin{figure}
\ContinuedFloat
\begin{center}
\includegraphics[width=0.75\linewidth]{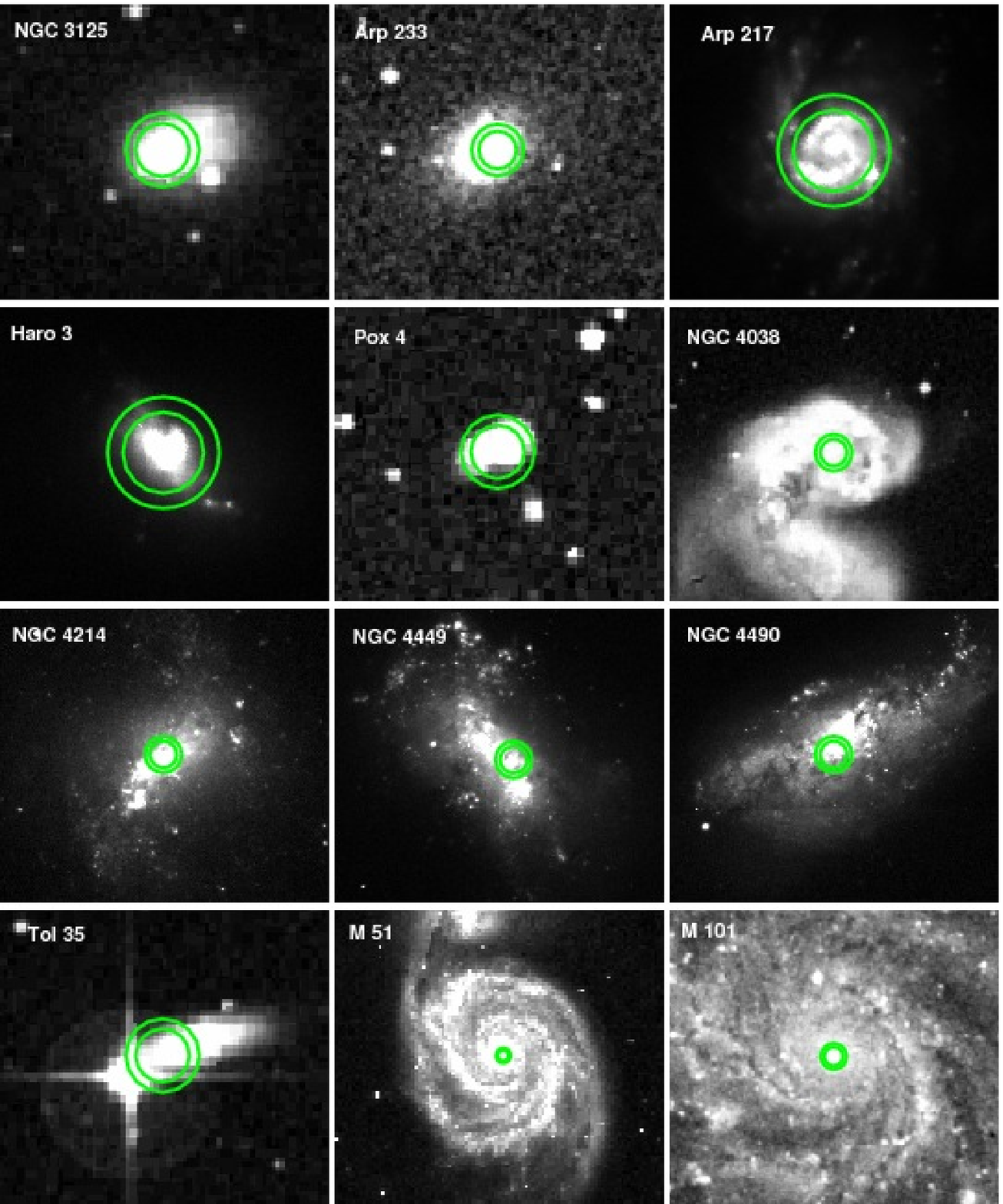}
\end{center}
\caption{Continued.}
\label{fig:beamsizes}
\end{figure}

\begin{figure}
\centering
\includegraphics[width=.3\textwidth, angle=270]{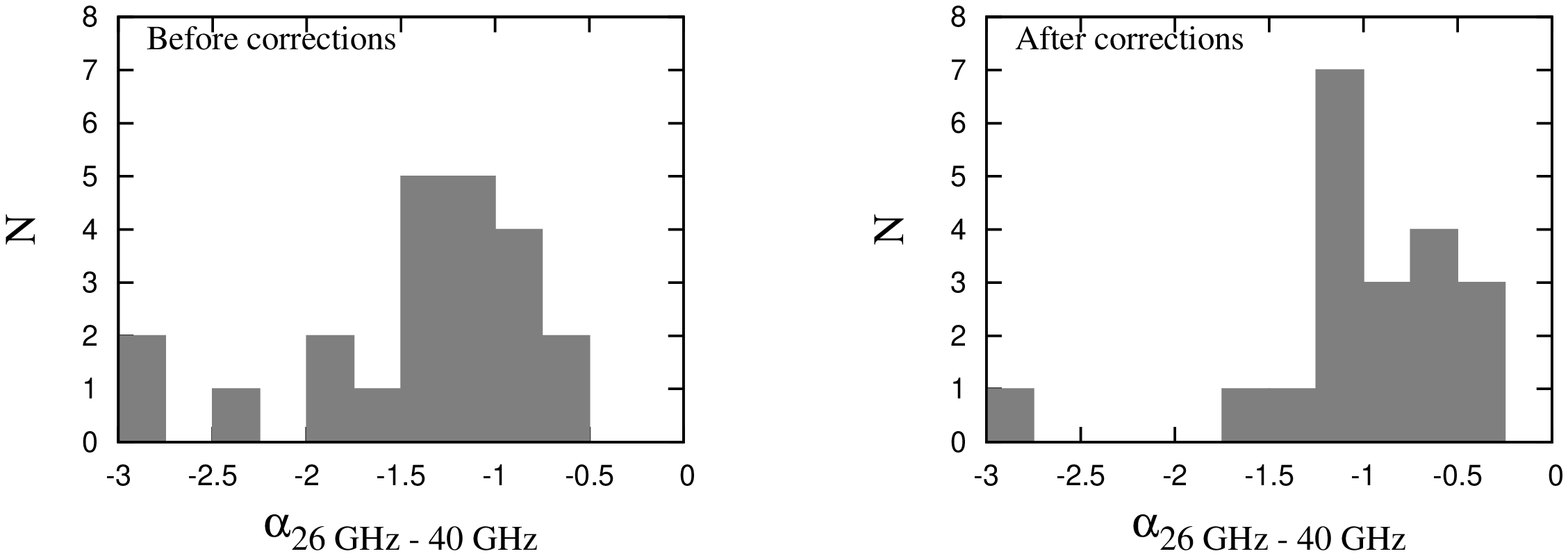}
\caption{Distribution of $\alpha_{26-40}$ before correcting for beam
  size (left) and after the corrections have been applied (right). The
  beam size corrections flatten $\alpha_{26-40}$ relative to
  uncorrected data.}
\label{fig:histogram}
\end{figure}

\begin{figure}
\centering
\includegraphics[width=15cm]{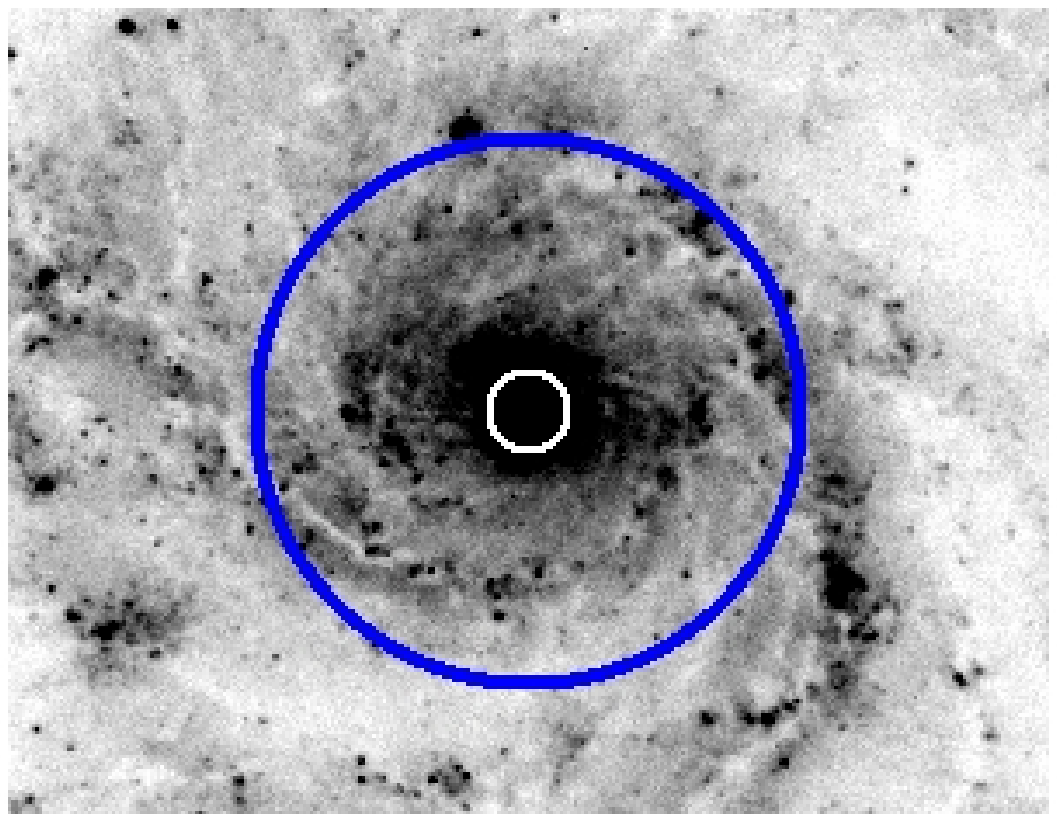}
\caption{Average CCB beamsize ($\sim 23\arcsec$, white) and circle
  with radius ($\sim 78\arcsec$) equal to the separation between the
  ``on'' and ``off'' beams (blue) overlaid on an optical (SDSS g)
  image of M 101. M 101 and M 51 are both larger than the beam
  separation, which likely results in an oversubtraction when the flux
  in the ``off'' beam is subtracted from the flux in the ``on'' beam.}
\label{fig:beamseparation}
\end{figure}

\begin{figure}
\centering
\includegraphics[width=0.75\linewidth]{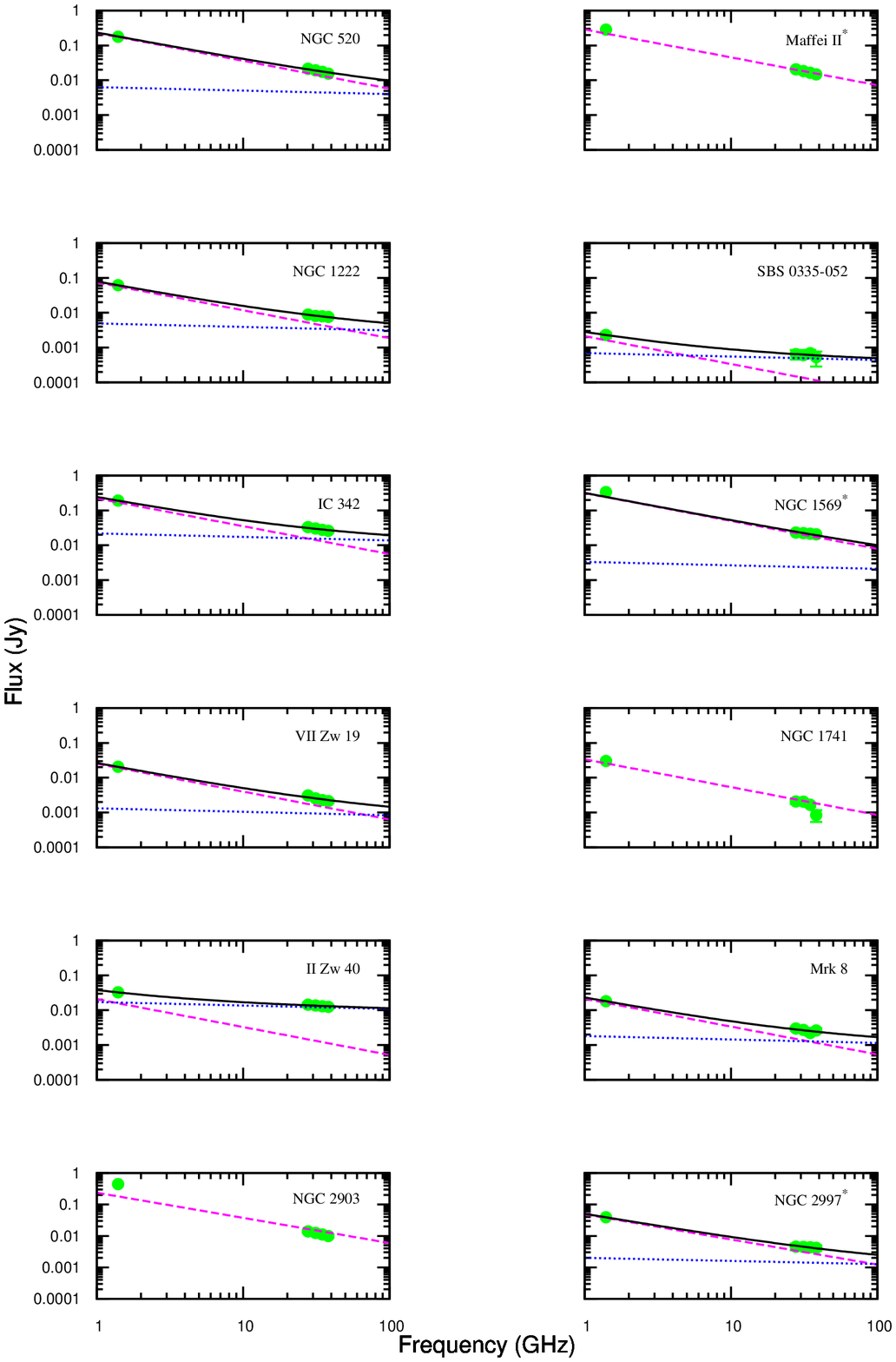}
\caption{\footnotesize NVSS 1.4 GHz and CCB 26-40 GHz points for each
  galaxy that was detected with all four CCB sub-bands. In
    most cases, the error bars are smaller than the point size. The
  best-fit spectral energy distribution for each galaxy is also
  plotted. Each SED was fit with a combination of nonthermal and
  thermal components (black line). The purple dashed line is the
  nonthermal ($\alpha_{N}=-0.8$) component, the blue dotted line is
  the thermal ($\alpha_{T}=-0.1$) component. When a galaxy's SED could
  not be fit with the inclusion of a positive thermal component, we
  only fit the nonthermal component (the thermal flux at 33 GHz is
  calculated as an upper limit in such cases). Galaxies that are
  resolved at 33 GHz are marked with an *.}
\label{fig:sedpanel}
\end{figure}

\begin{figure}
\ContinuedFloat
\centering
\includegraphics[width=0.75\linewidth]{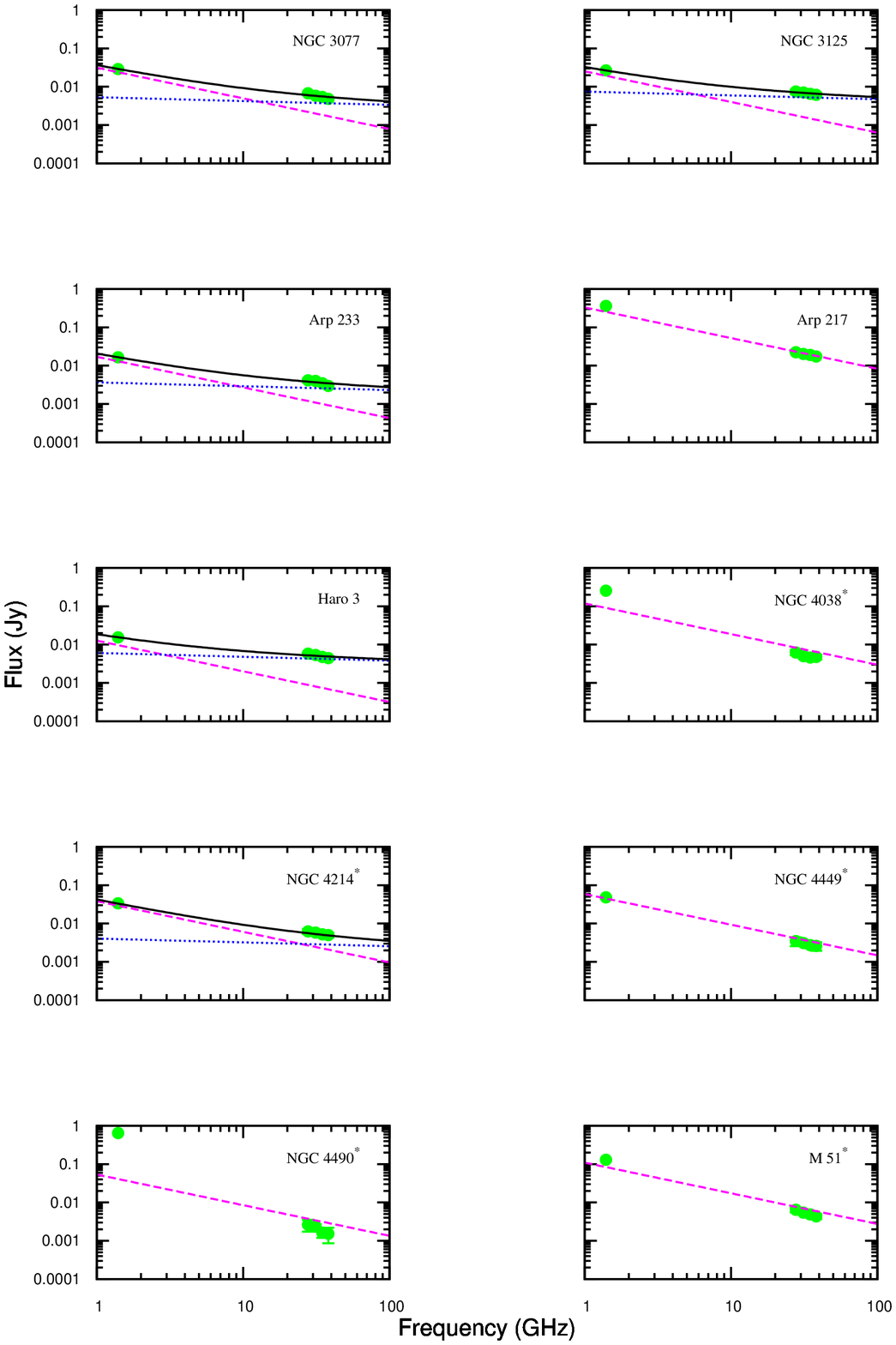}
\caption{Continued.}
\label{fig:sedpanel}
\end{figure}

\begin{figure}
\centering
\includegraphics[width=0.75\linewidth, angle=270]{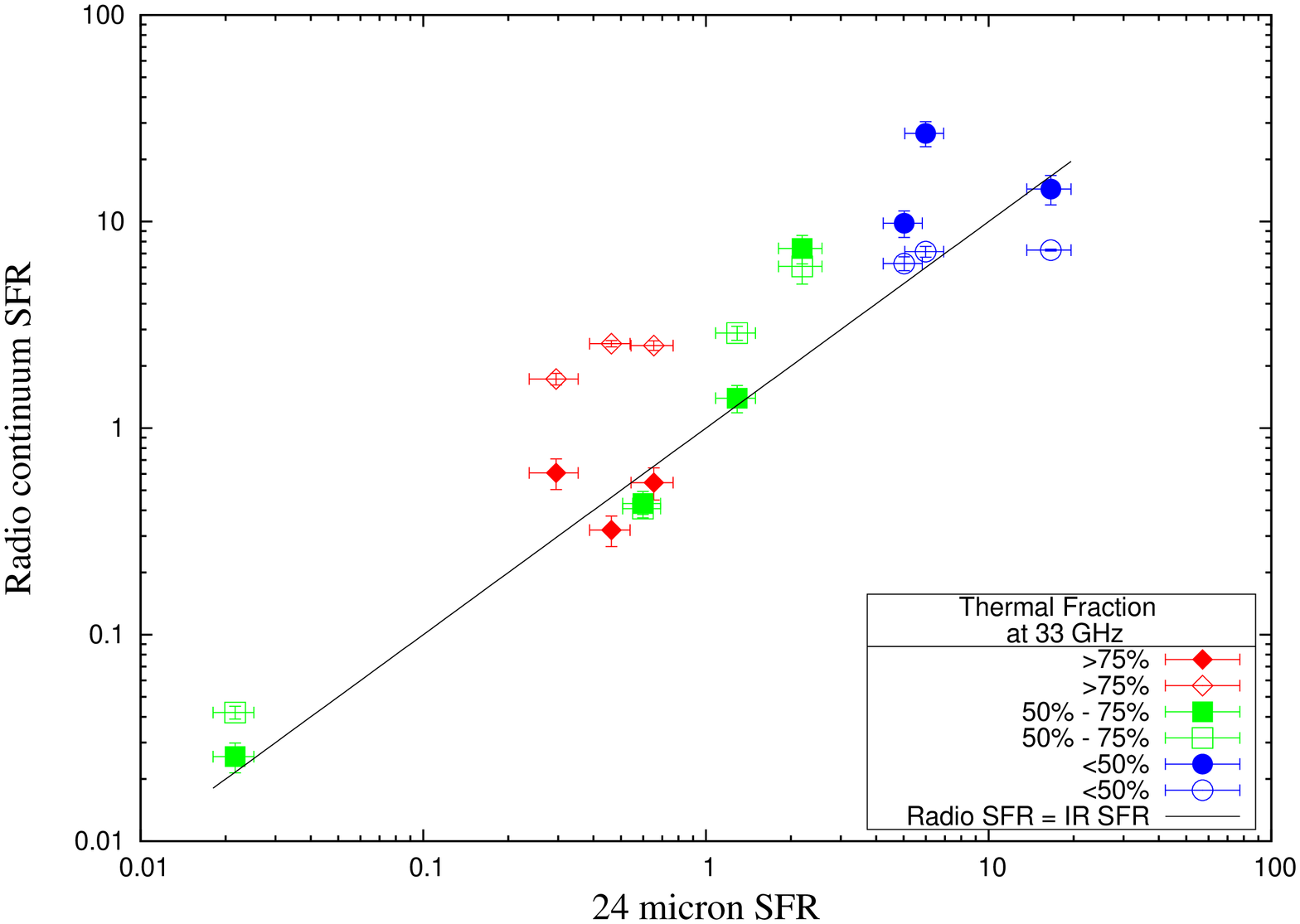}
\caption{Star formation rates calculated from nonthermal (filled
  symbols) and thermal (open symbols) radio continuum fluxes plotted
  against SFRs calculated from IRAS 25$\mu m$ fluxes according to
  Equations 1 and 17 from \citet{Cal10}. The solid black line represents
  equal SFRs at radio and infrared wavelengths. Most galaxies have
  higher SFRs when calculated using radio continuum fluxes, which
  trace more recent star formation than infrared fluxes.}
\label{fig:comparesfrs}
\end{figure}

\begin{figure}
\centering
\includegraphics[width=0.7\linewidth]{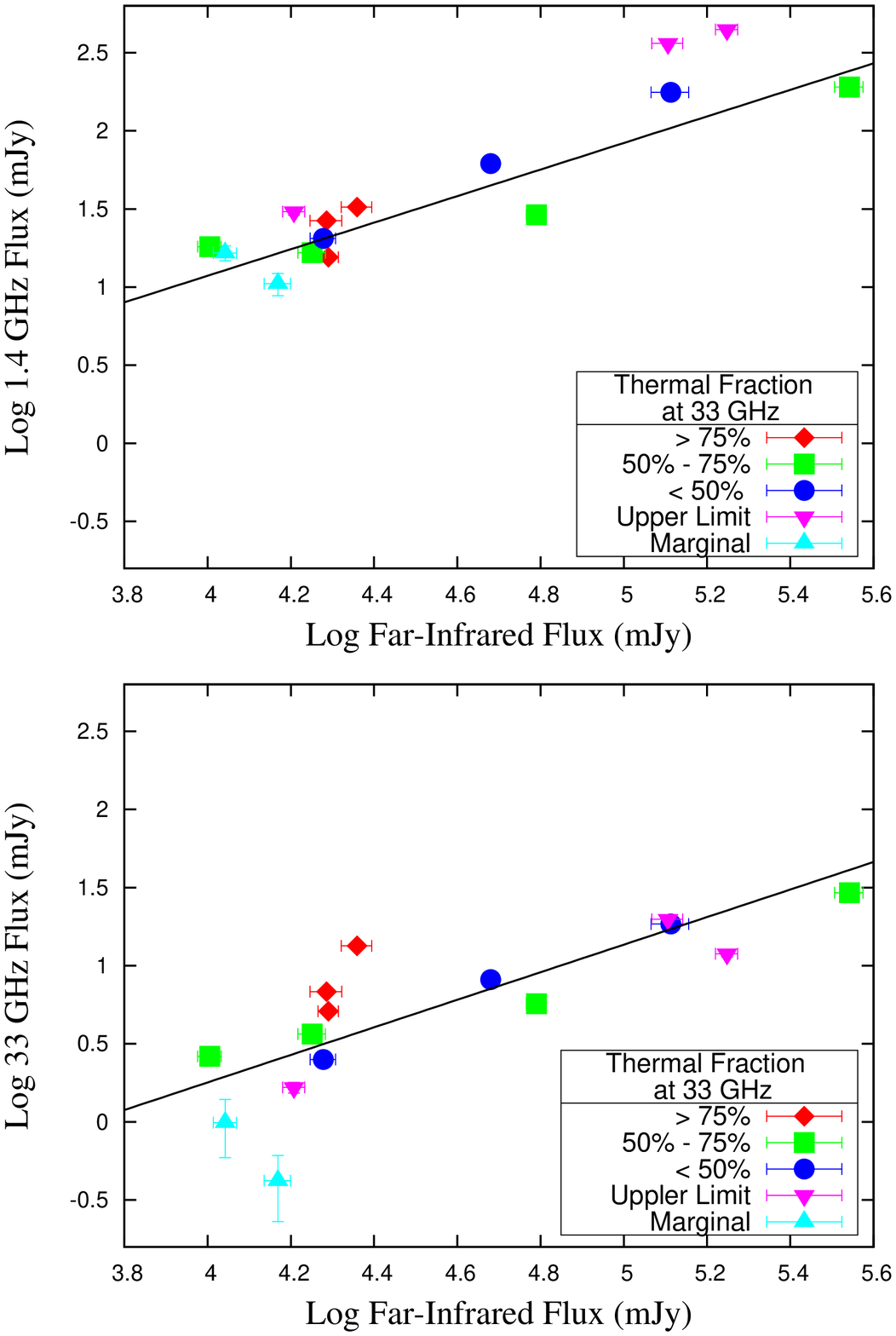}
\caption{\footnotesize Radio-far-infrared correlation for 1.4 GHz
  (top) and 33 GHz (bottom) fluxes vs. total far-infrared flux. The
  FIR flux is derived from a combination of IRAS 60 $\mu m$ and 100
  $\mu m$ data. All of the galaxies unresolved with the GBT's
  23$\arcsec$ beam are plotted except for SBS 0335-052, which was not
  detected by IRAS, and Pox 4, which was not detected at 100 $\mu
  m$. The lines that best fit each data set are also plotted. The
  galaxies are coded by thermal fraction at 33 GHz. While the
  correlation is tighter at 1.4 GHz than it is at 33 GHz, it is still
  easily seen at 33 GHz. Since the galaxies with the highest thermal
  fractions all lie above the best-fit line at 33 GHz (but don't at
  1.4 GHz), it is possible that some of the scatter in the correlation
  at 33 GHz is due to the increased proportion of thermal emission at
  higher frequencies.}
\label{fig:rad-IR}
\end{figure}

\begin{figure}
\centering
\includegraphics[width=0.7\linewidth, angle=270]{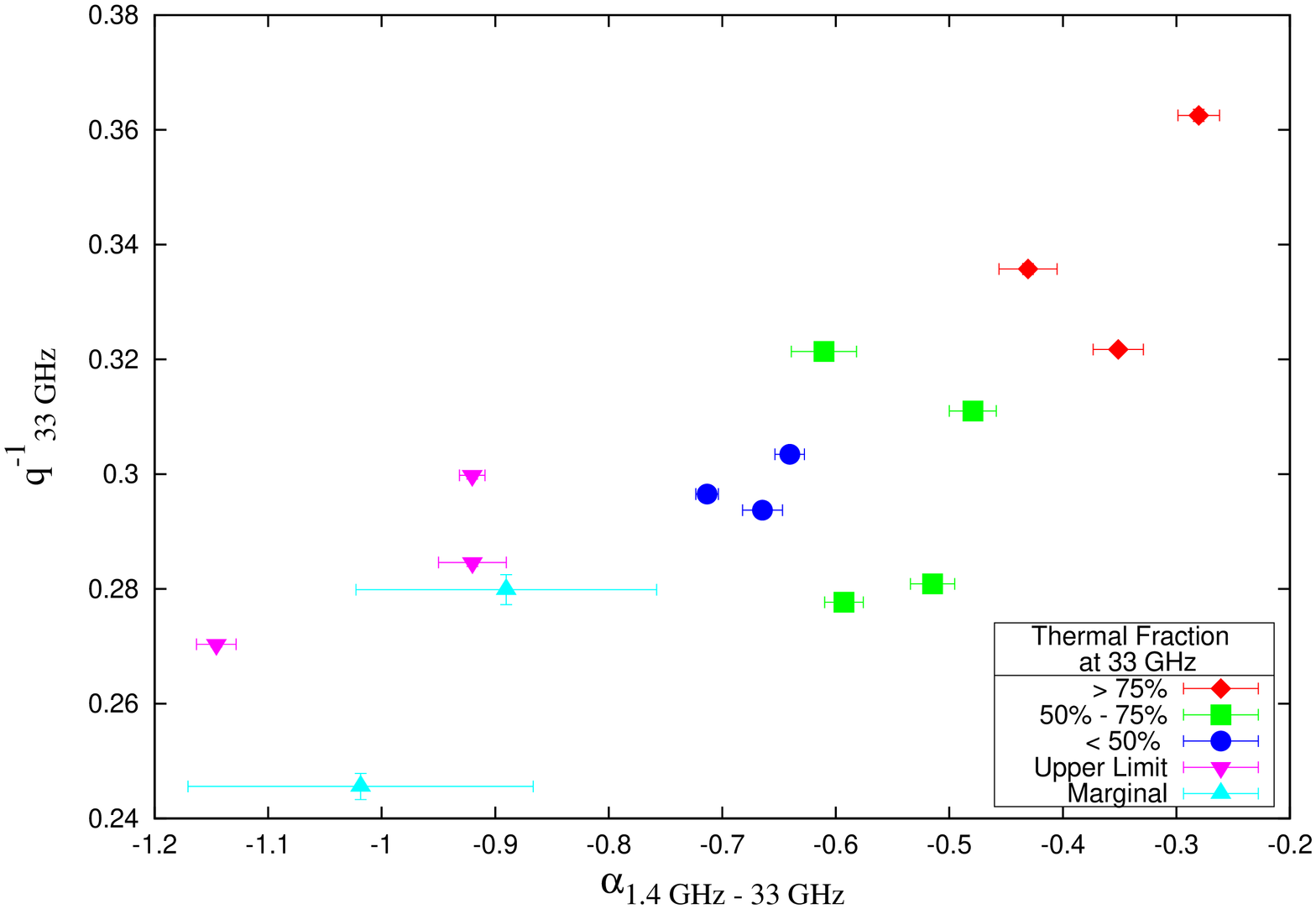}
\caption{$q_{33}^{-1}$ vs $\alpha_{1.4-33}$ using 33 GHz fluxes for
  unresolved galaxies. $q_{\nu}^{-1}$ is a measure of the ratio between
  radio flux and total far-infrared flux at a given radio
  frequency. As in Figure \ref{fig:rad-IR}, SBS 0335-052 and Pox 4 are
  not plotted. The red diamonds represent the highest thermal fraction
  (greater than 75$\%$). The green squares represent galaxies with
  thermal fractions between 50$\%$ and 75$\%$. The blue circles
  represent galaxies with thermal fractions less than 50$\%$. The
  purple inverted triangles represent galaxies where we were only able
  to determine upper limits for their thermal fractions. The light
  blue triangles represent galaxies that were only marginally detected
  at 33 GHz, so no thermal fraction was calculated. At 33 GHz, the
  ratio of radio flux to total FIR flux is highest when
  $\alpha_{1.4-33}$ is flat and thermal fractions are high. These
  three properties are all indicative of recent star formation. Thus,
  it is possible that these properties together act as a rough measure
  of the timescale of the current episode of star formation.}
\label{fig:inverseqvsalpha}
\end{figure}

\begin{figure}
\centering
\includegraphics[angle=270, width=\textwidth]{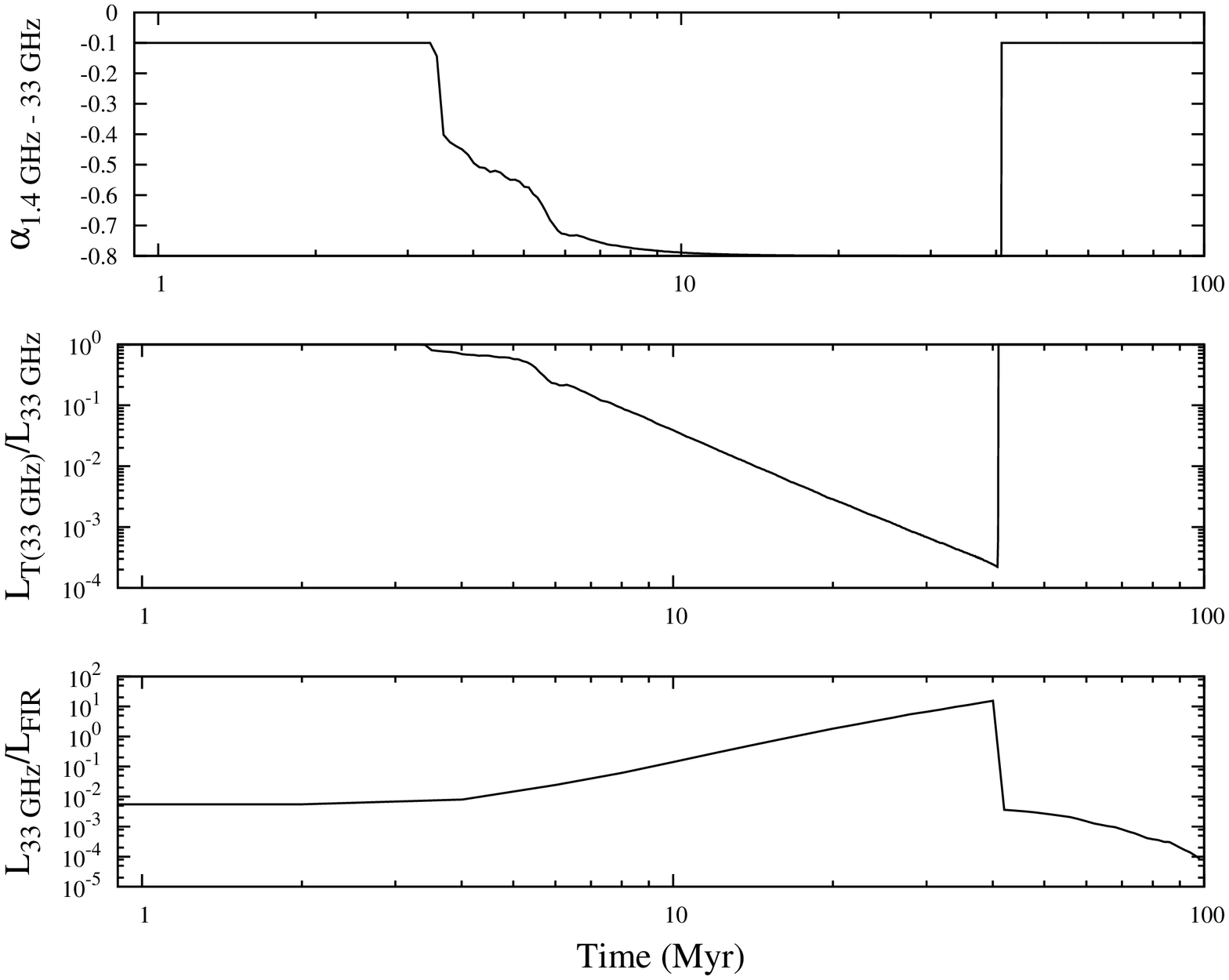}
\caption{\footnotesize Spectral index between 1.4 GHz and 33 GHz
  (top), thermal fraction (middle), and ratio of 33 GHz luminosity to
  total far-infrared luminosity (bottom) for a simple Starburst 99
  model of an instantaneous starburst. The large jumps in each curve
  at 40 Myr are due to the supernova rate dropping to zero at that
  time, as all of the stars massive enough to produce supernovae (and
  thus nonthermal emission) have died. The flattest spectral indices
  and highest thermal fractions are seen at the earliest times after
  the beginning of the starburst (up to 3 Myr), while the steepest
  spectral indices and loweset thermal fractions are seen at later
  times as more supernovae occur, up to 40 Myr, after which the
  supernovae cease. Similarly, the higher ratios of 33 GHz luminosity
  to FIR luminosity were seen during the lifetimes of massive stars,
  while the lowest ratios of 33 GHz luminosity to FIR luminosity were
  seen after supernovae ended, though this trend is delayed with
  respect to the timelines in the top two panels. This model
  demonstrates that flat spectral indices, high thermal fractions, and
  elevated 33 GHz fluxes with respect to FIR fluxes are all indicative
  of very recent star formation.}
\label{fig:sb99}
\end{figure}

\end{document}